\documentclass[aps,prb, onecolumn, nofootinbib, 11pt]{revtex4-2}
\usepackage{amsmath,amssymb,graphicx,dcolumn,bm,hyperref,float}
\usepackage{microtype}
\hypersetup{ colorlinks=true, linkcolor=red, filecolor=red,  urlcolor=red, citecolor = red}
\usepackage{xcolor}
\usepackage{csquotes}
\usepackage{tikz}
\usepackage{amsthm}
\usepackage[caption=false]{subfig}

\graphicspath{{./figures/}}

\begin{document}


\title{Random geometry at an infinite-randomness fixed point}

\author{Akshat Pandey}
\email{akshatp@stanford.edu}
\affiliation{Department of Physics, Stanford University, Stanford, CA 94305, USA}
\author{Aditya Mahadevan}
\affiliation{Department of Physics, Stanford University, Stanford, CA 94305, USA}
\author{Aditya Cowsik}
\affiliation{Department of Physics, Stanford University, Stanford, CA 94305, USA}
\date{\today}

\begin{abstract}
We study the low-energy physics of the critical (2+1)-dimensional random transverse-field Ising model. The one-dimensional version of the model is a paradigmatic example of a system governed by an infinite-randomness fixed point, for which many results on the distributions of observables are known via an asymptotically exact renormalization group (RG) approach. In two dimensions, the same RG rules have been implemented numerically, and demonstrate a flow to infinite randomness. However, analytical understanding of the structure of this RG has remained elusive due to the development of geometrical structure in the graph of interacting spins. To understand the character of the fixed point, we consider the RG flow acting on a joint ensemble of graphs and couplings. We propose that the RG effectively occurs in two stages: (1) randomization of the interaction graph until it belongs to a certain ensemble of random triangulations of the plane, and (2) a flow of the distributions of couplings to infinite randomness while the graph ensemble remains invariant.  This picture is substantiated by a numerical RG in which one obtains a steady-state graph degree distribution and subsequently infinite-randomness scaling distributions of the couplings. Both of these aspects of the RG flow can be approximately reproduced in simplified analytical models.

\end{abstract}

\maketitle


\section{Introduction}

Quenched randomness in quantum many-body systems can have dramatic effects at low energies. While it is sometimes the case that a small amount of randomness is perturbatively irrelevant in the renormalization group (RG) sense on top of a pure fixed point --- in which case the long-distance, low-energy physics can be treated as that of a translationally invariant field theory --- more interesting possibilities exist in both theory and experiment. 

For concreteness, consider problems concerning quantum magnets on a lattice in the presence of random couplings. Often, either due to constraints on the stability of a pure fixed point to weak disorder or because the disorder strength in the bare lattice model is large enough for notions of perturbative irrelevance in a field theory to be useless~\cite{Harris, CCFS}, disorder will remain an important feature of the low-energy theory.  Under a heuristic RG flow, the strength of randomness in the couplings of a theory will then either saturate to some finite value, or keep increasing without bound. 

So called infinite-randomness fixed points (IRFPs) represent the interplay of quantum mechanics and quenched randomness at its most stark: because of domination by rare-region effects, there is an exponentially big difference between certain averaged and typical quantities, and conventional dynamical scaling $t \sim \ell^z$ is destroyed in favor of ``tunnelling'' scaling $\ln t \sim \ell^\psi$. The effects of randomness are apparent in both dynamics and thermodynamics, as the two are tied together in quantum mechanics. 

Experimentally, lightly doped semiconductors have proven to be the most natural place in which to look for regimes of this strong-disorder physics: the non-interacting Anderson insulator is unstable to the formation of local moments once Coulomb repulsion is included, and these moments then interact with each other via exponentially decaying antiferromagnetic exchange interactions, which are broadly distributed even in the bare Hamiltonian~\cite{BhattLee}. Highly disordered fixed points have been proposed via local-moment instabilities of spin liquid and valence bond solid states in the presence of quenched randomness~\cite{KimchiNahumSenthil, KimchiLee, LiuSandvik2018, LiuSandvik2020}. Dirty metals also host a finite concentration of local moments; there RKKY interactions can qualitatively change the physics due to the relatively long-ranged ($1/r^d$) decay and oscillating ferromagnetic/antiferromagnetic sign of the couplings~\cite{FisherBhatt}.

The most powerful tool in the study of these random magnets has proven to be the strong-disorder renormalization group (SDRG)~\cite{IgloiMonthus, IgloiMonthus2}, first proposed by Ma, Dasgupta and Hu~\cite{MDH}, and further elaborated on by D.S. Fisher~\cite{fisher1994random, fisher1995critical}. The central premise of the SDRG is that a broad disorder distribution naturally gives rise to a small parameter, namely the ratio of a typical term in the Hamiltonian to the largest term. Doing perturbation theory in this small parameter allows one to integrate out the strongest term in the Hamiltonian and to renormalize the couplings geometrically adjacent to this term in the lattice. The protocol of SDRG is to repeatedly integrate out the largest term in the Hamiltonian, each time modifying the remaining terms and thereby defining a flow of the Hamiltonian with decreasing energy scale. If, starting with a broad disorder distribution, the RG makes the distribution grow broader without bound --- i.e. if it takes us to an IRFP --- then the approximation of the largest term being much stronger than the terms neighboring it can be justified self-consistently, and the RG flow becomes asymptotically exact at low energies.

A simple example of a model that flows to infinite randomness is the (1+1)-dimensional random transverse-field Ising model (TFIM),
\begin{equation}\label{eq:H1d}
    H = - \sum_i J_i \sigma_i^z \sigma_{i+1}^z -\sum_i h_i \sigma_i^x,
\end{equation}
at its ferromagnet-to-paramagnet critical point, which occurs when the distributions of logarithms of the $J_i$ and $h_i$ have equal means: $\langle \ln J \rangle = \langle {\ln h} \rangle$ (we assume that the $J_i$ and $h_i$ are independent and identically distributed, each according to its respective distribution). The SDRG theory of this phase transition, and of the randomness-induced critical phases next to it, was developed in Ref.~\cite{fisher1995critical}.
\begin{figure}
    \centering
    \includegraphics[width=0.6\textwidth]{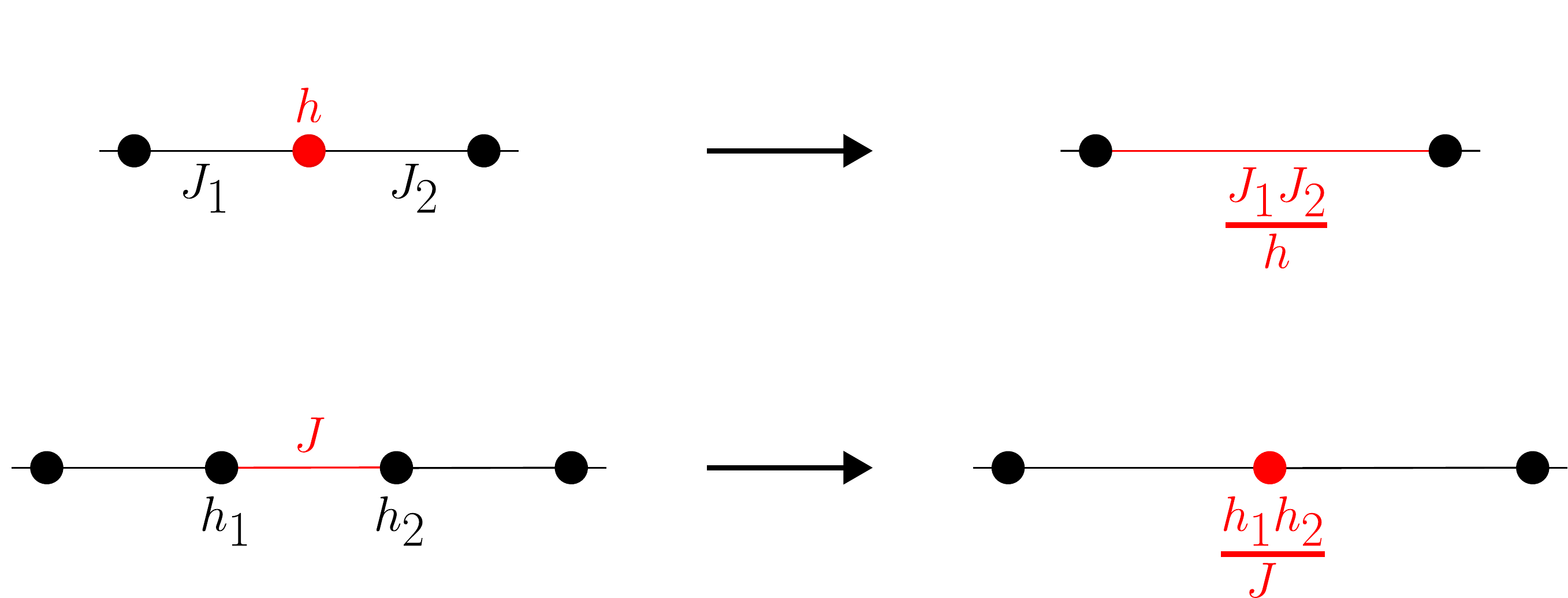}
    \caption{Decimation rules for strong-disorder RG in the (1+1)d transverse-field Ising model. The upper and lower subfigures depict decimations when the strongest coupling on the chain is a transverse field $h$ and an Ising bond $J$ respectively. Note that, under these RG steps, the chain geometry does not change, apart from a trivial reduction by one lattice site. }
    \label{fig:1drules}
\end{figure}

The RG rules are summarized in Fig.~\ref{fig:1drules}. If a spin is decimated it is frozen into the $+x$ direction and a new bond is made between the spins formerly adjacent to it. If a bond is decimated the two spins that it connected become one big spin, and the bond goes away. Crucially, under these RG rules the geometry of the interaction graph is unchanged from the 1d chain and the remaining couplings and fields on the chain remain uncorrelated, so the RG flow can be described in terms of the flow of the bond and field distributions under successive decimations.

Consider implementing a generalization of the same rules in two dimensions starting from some lattice (again with all $J_i$ as well as all $h_i$ independent and identically distributed to begin with). Now, when a spin with transverse field $h$ which has $n$ neighbors (``degree'' $n$ in graph theoretic  language) is decimated, in second-order perturbation theory one has to remove $n$ bonds, $J_1, J_2, \dots, J_n$, and make $n(n-1)/2$ new ones with strengths $J_i J_j /h$. This  immediately modifies the interaction graph. If one starts with a lattice, the new graph is no longer a lattice, and in fact tends to develop bigger and bigger all-to-all cliques which make the graph grow closer to complete. The model that is closed under the action of the RG should hence be defined on general graphs $G = (V, E)$ as
\begin{equation}
    H = -\sum_{\langle ij\rangle \in E} J_{ij} \sigma_i^z \sigma_j^z - \sum_{i \in V} h_i \sigma_i^x .
\end{equation}
We stress that we will study disorder distributions with all $J_{ij} > 0 $, i.e. ferromagnetic, and all $h_i > 0$ as well. Including both signs of the couplings makes the problem more complicated, although it has been conjectured that this modification is irrelevant near infinite randomness~\cite{motrunich2000infinite}. 

The other ingredient of the 2d problem which does not exist in the 1d one is the existence of correlations between couplings. For example, two bonds created during the RG may have a constituent bond in common, as in $J_1J_2/h$ and $J_1 J_3/h$. There will also be correlations between geometrical quantities and couplings --- for example spins with small transverse fields will tend to have big degrees because it took a large number of decimations to make them.

These issues constitute qualitative barriers to analytical considerations, and impede a complete understanding of the SDRG compared to studies in one dimension. Nevertheless, the RG rules have been numerically implemented~\cite{motrunich2000infinite}, and the quantum critical point is found to flow to infinite randomness, as in 1d. Critical exponents that determine thermodynamics and disorder-averaged correlation functions can be computed in this numerical RG. 

It follows from the above discussion that, in going from one to two dimensions, one needs to generalize the notion of the SDRG from a flow of self-similar couplings on a fixed geometry (viz. the chain) to a potentially joint flow of self-similar couplings \emph{together with the geometry}.

The central object of our study is a geometrically motivated truncation of the full RG rules, shown in Fig.~\ref{fig:triangrules}. Here, when a field $h$ on a spin with degree $n$ is decimated, one orders the bonds next to it as $J_1 > J_2 > \dots > J_n$, and only makes new bonds with magnitudes $J_1 J_i /h$. We conjecture that this truncation is irrelevant at the IRFP, and this is verified by implementing the RG numerically. We shall argue for validity of our truncation and explain how infinite randomness in the couplings is intertwined with the geometrical ensemble that appears at the fixed point of our RG. 

Triangulations of the plane (or of the torus in numerics with periodic boundary conditions) are almost preserved by our RG, modulo subtleties which will be discussed presently. Furthermore, the RG makes the effects of site and bond decimations on the interaction graph identical, as can be seen in Fig.~\ref{fig:triangrules}. The geometry thus generated is naturally much simpler than the geometry in the  RG without truncations, i.e. the one that generates all bonds that can be generated in second-order perturbation theory ~\cite{motrunich2000infinite}.

In fact the triangular truncation reveals the structure of the RG flow rather vividly. We find that the flow occurs in two steps: (1) Starting from the bare lattice geometry and bare distributions of couplings, the first step is mostly geometrical. The geometry quickly approaches a steady state, a ``fixed-point geometry,'' which is approximated by a certain ensemble of random triangulations. The couplings during this stage feature primarily in how they feed into the evolution of the geometry from the initial lattice to the fixed-point geometry. (2) In the second stage, the geometry wanders ergodically through the fixed-point ensemble, while the couplings flow to infinite randomness on top of it.

Both of these stages can be approximated in analytical toy models (once we ignore a variety of correlations). The most important geometrical quantity is the distribution of degrees $n$ of the graph, which we will call $p(n)$. In the first stage, we can write down a differential equation for the evolution of $p(n)$ with RG time and find its steady state, having fed in a caricature of the information about correlations between degrees and coupling strengths which determine the rates at which the geometrical operations occur as a function of $n$. All further correlations, which in principle exist, are dropped. The second stage involves coupled integro-differential equations similar to those analyzed by Fisher~\cite{fisher1994random}, for the evolution of the distributions of couplings $J$, $h$. These differential equations take as input the fixed-point geometry's degree distribution $p(n)$, and again ignore all correlations beyond those which have to be assumed to maintain the geometrical steady state. With $p(n)$ chosen to be of a functional form that greatly simplifies the RG equations (and is a good fit to the numerically observed $p(n)$) we find an approximate scaling solution which flows to infinite randomness, in fact with critical exponents that are close to the numerically observed ones.

The remainder of the paper is structured as follows. Section~\ref{sec:background} contains background on the problem (in both the one- and two-dimensional cases). In Section~\ref{sec:triangular} we introduce the RG rule, present numerical results obtained using it for infinite-randomness scaling and the evolution of geometry, and interpret these results using the two-stage perspective described above. In Section~\ref{sec:analytical} we outline solutions to analytical toy models for the RG. Finally Section~\ref{sec:discussion} concludes with a discussion of open problems.

\section{Background}\label{sec:background}

In this section, we mostly explain previous work and set up notation.

\subsection{One dimension}

Here we discuss in more detail the random one-dimensional TFIM, Eq.~\eqref{eq:H1d}, following Ref.~\cite{fisher1995critical}. One can obtain the ground-state and low-energy properties of this Hamiltonian using the following approximate RG, summarized in Fig.~\ref{fig:1drules}.

Find the largest of the set of couplings, $  \Omega = \max_i\{ J_i, h_i\}$. If $\Omega$ corresponds to a bond $J_i$, remove the bond, and combine the two spins it connects to form one new spin with a larger moment and is acted on by a new transverse field $ h_ih_{i+1}/J_i$. If $\Omega$ corresponds to a transverse field $h_i$, put the site in its ground state (pointing along $x$), remove it as well as the two bonds $J_{i-1}$ and $J_i$, and add in a new renormalized bond connecting the sites which were previously called $i-1$ and $i+1$ with strength $J_{i-1}J_{i+1}/h_i$. These RG steps correspond to integrating out the strongest coupling and creating new couplings to second order in perturbation theory. The ratio of the biggest coupling to those next to it is always treated as a large parameter, and so the approximation is sensible only if the distributions of $J$'s and $h$'s are broad. Indeed an eventual RG fixed point is controlled only if the RG is asymptotically exact, in other words if it makes the distributions flow to be infinitely broad.

It is convenient to work in logarithmic variables, which will always be nonnegative:
\begin{equation}
    \zeta = \ln (\Omega/J), \quad \beta = \ln (\Omega/h), \quad  \Gamma = \ln(\Omega_0/\Omega).
\end{equation}
With $\Omega_0$ equal to the initial largest scale, $\Gamma$ is an RG time. In terms of these logarithmic variables, the RG steps will be additive, and the denominators will play no role. The objects that flow under RG are the distributions of the log couplings $P(\zeta)$ and $R(\beta)$:
\begin{equation}\label{eq:1dRGP}
    \frac{\partial P }{\partial \Gamma}= \frac{\partial P}{\partial \zeta} + R_0\int d \zeta_1 \,  P(\zeta_1) P(\zeta - \zeta_1) - 2R_0 P + (P_0 + R_0)P,
\end{equation}
\begin{equation}\label{eq:1dRGR}
    \frac{\partial R }{\partial \Gamma}= \frac{\partial R}{\partial \beta} + P_0\int d \beta_1 \,  R(\beta_1) R(\beta - \beta_1) - 2P_0 R + (P_0 + R_0)R,
\end{equation}
where $P_0 = P(0)$ and $R_0 = R(0)$. We will explain \eqref{eq:1dRGP}; \eqref{eq:1dRGR} has an analogous form due to the field-bond duality in this model.

The first term comes from the fact that once the log energy shell of size $d\Gamma$ is integrated out, $\Omega$ is redefined to be the new strongest coupling so that the whole distribution shifts leftward by $d\Gamma$, or equivalently $P$ increases by $d\Gamma  \times \partial P/\partial \zeta$. The second term adds in all the newly made bonds of size $\zeta$ created by the decimation of a field. It receives contributions, therefore, from all bond-bond pairs whose log strengths sum up to $\zeta$. The third term describes the removal of these two bonds, which are sampled independently from the distribution. Both of these processes occur $d\Gamma\times R_0$ times, as this is the fraction of sites decimated. Finally, the last term keeps the distribution normalized: the integral of the right hand side with respect to $\zeta$ vanishes.

The simplest solution of these equations is
\begin{equation}
    P(\zeta) = \frac{1}{\Gamma} e^{-\zeta/\Gamma}, \quad R(\beta) = \frac{1}{\Gamma}  e^{-\beta/\Gamma},
\end{equation}
and this corresponds to the RG flow at the ferromagnet-paramagnet critical point. Clearly, the width of both distributions grows broader without bound as $\Gamma$ increases: in the original $J/h$ language, the \emph{orders of magnitude} spanned grow infinitely broad. This defines an infinite-randomness fixed point. 

A number of other properties of the system at and off criticality can be calculated using the SDRG; we will not discuss these here, and mostly will not in our study of the two-dimensional case either. The one exception is the number of active (i.e. non-decimated) spins in the cluster labelled by site $i$, or equivalently the magnetic moment $\mu_i$ of cluster $i$. Initially the magnetic moment of each spin is $1$, and when two spins with moments $\mu_1$ and $\mu_2$ are joined via a bond decimation, the moment of the new spin is $\mu_1+\mu_2$. Eq.~\eqref{eq:1dRGR} can be modifed to include this auxiliary variable, and the joint distribution for $ \beta$ and $ \mu$ can be solved for as well.

\subsection{Two dimensions}

It is straightforward to define RG rules for the TFIM on a general graph, of which the one-dimensional chain is the simplest case. Now, when a site carrying transverse field $h$ and with $n$ neighbors is decimated, we have to destroy the $n$ bonds that are connected to the decimated site --- call these $J_1, J_2, \dots, J_n$ --- and create all bonds $J_i J_j/h$ that are generated in second-order perturbation theory assuming $h/J_i \gg 1$ for all $i$. Some of these bonds may already exist in the graph, in which case one replaces a bond with the newly created one if the new bond is stronger than the old one. This approximation is justified in the limit of infinite randomness, because in this case one of the new/old bonds is expected to be much stronger than the other one. 

Bond decimations contract edges of the graph, renormalizing the strength of a transverse field on a cluster of two spins to $h_1 h_2/J$ where $h_1$ and $h_2$ were the transverse fields on the spins shared by the decimated bond $J$. Any bonds to common neighbors of the two spins being fused will, similarly to above, take on the value of the stronger of the two bonds that overlap after the decimation. 

A generic graph will be greatly modified by these RG rules. In particular, lattices in two dimensions (which constitute the graph ``initial conditions'' for the RG trajectories we are interested in) will lose their lattice character immediately. These RG rules were first implemented numerically in Ref.~\cite{motrunich2000infinite} on two-dimensional lattices, keeping track of the interaction graph as the RG progresses. The log coupling distributions $P(\zeta), R(\beta)$ broaden without bound and reach a scaling form at criticality, thus establishing the existence of an IRFP in two dimensions. 

Since the RG is not on as firm analytical ground in two dimensions as in one, it should be checked against unbiased numerics, in particular against quantum Monte Carlo which can be performed on this model without a sign problem. The Monte Carlo numerics also shows an IRFP, with critical exponents well within error bars compared to the RG~\cite{Pich, KangSuperuniversality}. Therefore we will take as given the existence of the IRFP and rough values of its critical exponents.

We do not intend to provide a full review of SDRG studies of the TFIM (see Refs.~\cite{IgloiMonthus, IgloiMonthus2} for a guide to the SDRG literature on this and related systems) but would like to mention two approaches to modified SDRG schemes which drop a number of bonds produced by spin decimations and have strongly influenced this work.

Kov\'acs and Igl\'oi in Refs.~\cite{KovacsIgloi, KovacsIgloi3d, KovacsIgloiJPhys} determine a number of rigorous truncations of the RG via which the creation of the all-to-all cliques alluded to above can be avoided while leaving the RG trajectory effectively unchanged (at long enough RG times). These truncations (along with other optimizations) turn out to be extremely powerful, and allow numerics on systems with $ N \sim 2000^2$ spins. The effectiveness of these truncations also implies that a great number of the bonds that seem to make the graph look locally all-to-all are actually irrelevant at the fixed point, and need not be treated on an equal footing with the other bonds. Of particular note is the site decimation rule in Ref.~\cite{KovacsIgloi3d}, whose action is very similar to ours. Our rule is simpler in comparison and makes some analytical approximations possible, at the cost of rigor in demonstrating that the truncation is irrelevant at the IRFP.

A less rigorous truncation was proposed in Ref.~\cite{laumann2012strong} by Laumann \textit{et al}. These authors preserve planarity of the graph in a greedy manner: when a degree-$n$ site is decimated, the tentative new bonds are added from strongest to weakest until one encounters a planarity-breaking bond, at which point one ignores this bond and proceeds to the next strongest one, until all $n(n-1)/2$ bonds have been considered. This scheme is correct in that it flows to an IRFP with the same critical exponents, and also speeds up numerics. 

Most importantly, the success of these schemes lead one to the geometrical principle that strong non-planarity is not present at the two-dimensional IRFP. This is \emph{a posteriori} natural, as planarity is the graph theoretic measure of two dimensionality. Triangulations of the plane hold a special position among planar graphs as those in which the most bonds are retained: we will define what this means presently.

Here we define the critical exponents of the IRFP, which allow us to numerically check that we are at the correct fixed point. IRFPs display so-called tunneling dynamical scaling so that lengths scale as powers of \emph{logarithms} of times, or equivalently logarithms of energies:
\begin{equation}
  \langle \beta \rangle  \sim \langle \zeta \rangle   \propto N^{-\psi/d},
\end{equation}
where $N$ is the number of undecimated spins in the graph at some RG time. This defines the tunneling dynamical critical exponent $\psi$. We also measure a ``fractal dimension'' $d_f$ for the spin clusters, defined as
\begin{equation}
    \langle \mu \rangle \propto N^{-d_f/d}.
\end{equation}
Henceforth we will only work with $d = 2$. Definitions of several of the  quantities introduced above are summarized in Table \ref{table:notation}.

\begin{table}[H]
\centering
\begin{tabular}{c | @{\hskip .5cm} l}\label{table:definitions}
$\Omega$ &  $\max_{ij} \{ J_{ij}, h_i\}$,  largest coupling \\\hline
$\Omega_0$ &  $\Omega$ for bare Hamiltonian \\\hline
$\Gamma$ & $\ln(\Omega_0/\Omega)$, RG time \\\hline
$\zeta_{ij}$ & $\ln(\Omega/J_{ij})$, log bond  \\\hline
$\beta_i$ & $\ln(\Omega/h_i)$, log transverse field\\\hline
$\mu_i$ & number of active spins in vertex $i$ \\\hline
$n_i$ & degree of vertex $i$\\\hline
$P(\zeta)$ & probability distribution of log bonds \\\hline
$R(\beta)$ & probability distribution of log fields \\\hline
$p(n)$ & probability distribution of degrees \\\hline
$N$ & number of remaining spins \\\hline
$\psi$ & tunneling exponent, $\langle \beta \rangle \sim \langle \zeta \rangle  \propto N^{-\psi/d}$ \\\hline
$d_f$ & fractal dimension exponent, $\langle \mu \rangle \propto N^{-d_f/d}$

\end{tabular}

\caption{Definitions of commonly used quantities.}
\label{table:notation}
\end{table}

\section{Numerical results}~\label{sec:triangular}

\begin{figure}
    \centering
    \includegraphics[width=0.6\textwidth]{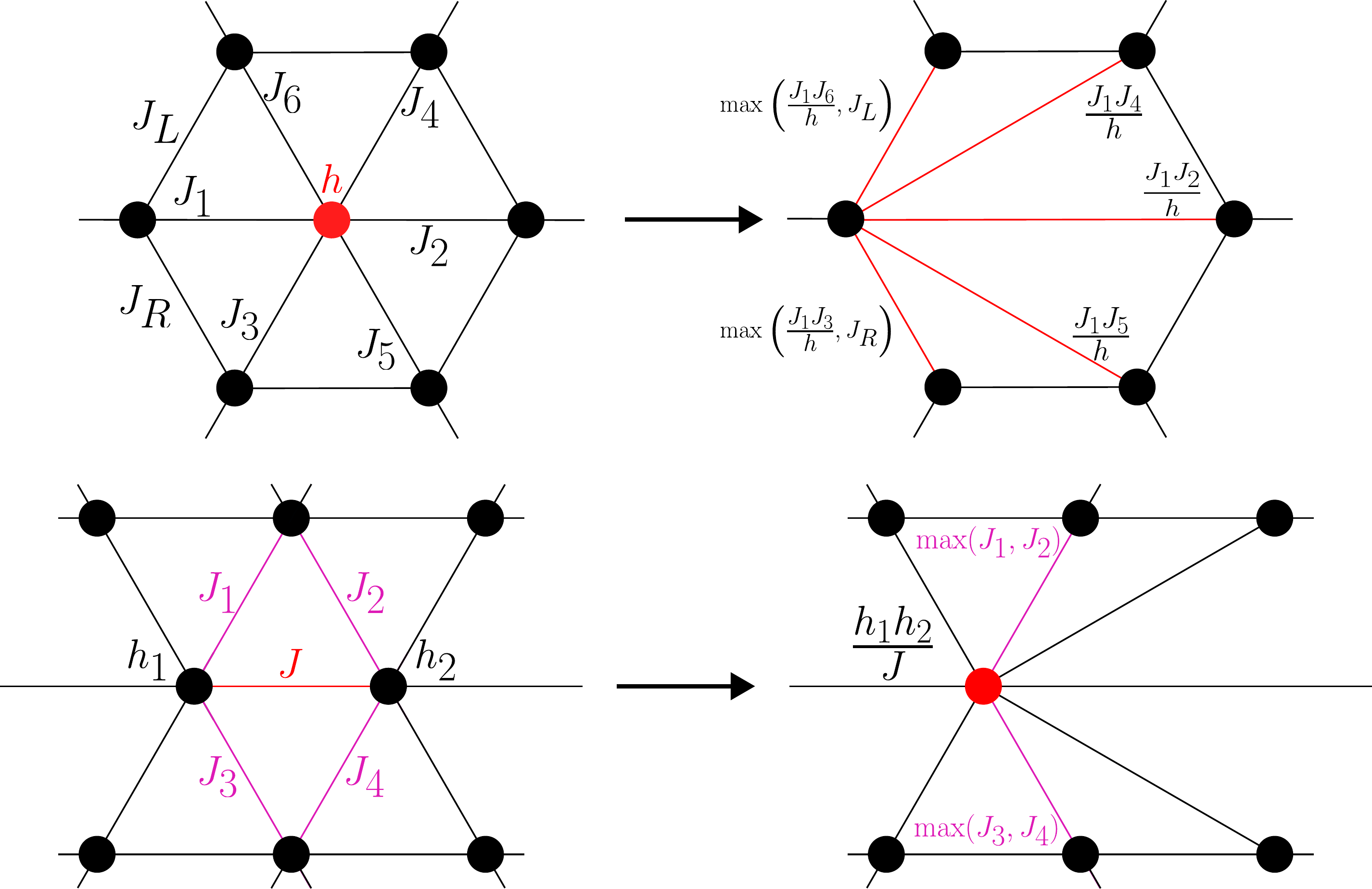}
    \caption{An example of a site/bond decimation (upper/lower subfigures respectively) according to our triangular RG rule in the (2+1)d transverse-field Ising model. For site decimation the order of adjacent bond couplings $(J_1 > J_2 > \dots)$ identifies the spin that the triangulating bonds all emerge from. Note that the geometrical modification enacted on the graph by these two decimations, once other neighboring bonds are drawn, is identical.
    }  
    \label{fig:triangrules}
\end{figure}

\subsection{RG rules}
The two-dimensional RG rules discussed in the previous section are of varying degrees of complexity in how many bonds they effectively truncate and what families of graphs they leave invariant. However, none of them seem amenable to analytical treatment due to the way in which they mix the geometrical and coupling subsets of the problem. To get around this issue, we discuss an RG rule which maintains planarity but is simpler in that it lends itself to a geometric mean-field type of analysis.

One simple way to maintain planarity of a graph is to only perform bond decimations, which contract edges of the graph. Therefore we propose a rule for spin decimations which looks, from the perspective of graph connectivity, exactly the same as a bond decimation. If decimating a field on spin $i$, this involves redirecting all bonds in which $i$ is involved to one of its neighbors $j$ (cf. Ref.~\cite{KovacsIgloi3d}). In order to bias toward keeping larger bonds, we choose $j$ to be the spin bound most strongly to $i$. 

Bond decimation does not create more edges in the graph, and does not break planarity, and we can implement it as usual, merging the two spins connected by the decimated bond and renormalizing the field on the cluster spin. 

The decimation rules are illustrated in Fig.~\ref{fig:triangrules}.

To summarize the rules for the triangular RG, if the strongest term in the Hamiltonian is the bond $J_{ij}$, between spins $i$ and $j$:
\begin{enumerate}
    \item For all $k \neq i$ adjacent to $j$ remove $J_{jk}$ and make bonds between $i$ and $k$ of strength $\max \left(J_{ik}, J_{jk}\right)$ where the value of a nonexistent bond is taken to be $0$. 
    \item Set the new field on $i$ to $h_i h_j / J_{ij}$.
    \item Remove $J_{ij}$ and the spin $j$.
\end{enumerate}

Instead if the strongest term in the Hamiltonian is the field $h_i$ on spin $i$:
\begin{enumerate}
    \item Find $j$, the spin adjacent to $i$ such that $J_{ij} > J_{ik}$ for all $k \neq j$ also adjacent to $i$. 
    \item Make bonds between $j$ and $k$, for all $k \neq j$ adjacent to $i$, of value $\max\left(J_{jk}, J_{ij}J_{ik}/h_i \right)$. 
    \item Remove $i$ and all bonds $J_{il}.$
\end{enumerate}

It can also be seen that these decimations very nearly map one triangulation of the plane to another. We will discuss caveats to this statement, but the fixed-point geometry will eventually be characterized as random planar triangulations in our analysis.

For numerical implementation, storing the terms of the Hamiltonian (fields and bonds) in a priority queue makes retrieving the largest term efficient, and we pop terms from the queue as they are decimated. When bonds are superseded by stronger bonds formed, or a field is renormalized, we can deactivate the old bond or field which is more efficient than removal from the priority queue.

We implement the numerics on a triangular lattice with periodic boundary conditions, so that the graph is planar on a torus. We have additionally run numerics on a square lattice to see that our results do not change: this will be discussed later in the context of the relation between renormalized lattice geometry and couplings. Throughout the paper, numerical results are presented from a $500\times 500$ lattice, and averaged over $100$ independent disorder realizations.

\subsection{Infinite-randomness scaling}

In order to check the validity of our numerical RG at criticality, we look for the existence of an infinite-randomness critical point for some initial distribution of disorder. At infinite randomness, $P(\zeta)$ and $R(\beta)$ scale in the same way, as all the energy scales in the problem flow together. However, in the paramagnetic phase $P(\zeta)$ broadens faster than $R(\beta)$, meaning that the fields predominate at low energies. Conversely in the ferromagnetic phase $R(\beta)$ broadens faster than $P(\zeta)$. Therefore one way to diagnose an IRFP is to measure the mean $\beta$ and $\zeta$ and find a distribution of initial couplings and fields such that the ratio $\langle\beta\rangle/\langle\zeta\rangle$ reaches a constant in the scaling regime of the RG. Since $R(\beta)$ remains roughly exponential throughout the RG and $P(\zeta)$ develops a hump at nonzero $\zeta$ (as shown in, e.g., Ref.~\cite{motrunich2000infinite}), we start with initial distributions of this form to minimize transients (though without the essential correlations among couplings and between the couplings and the graph that develop during the RG). 

Fixing $R(\beta) = e^{-\beta}$ (i.e. exponential with unit mean) at the start of the RG in all our numerics, we parametrize the initial $\zeta$ distribution, similarly to Ref.~\cite{laumann2012strong}, by $P(\zeta)= \frac13(a+b\zeta)e^{-c\zeta}$ and define the quantity $m=b-ac$, so that $m/3$ is the slope of the distribution at $\zeta=0$ and $a/3$ is the intercept. The one-parameter family of initial distributions that we use to go through criticality is generated by fixing $a$ and varying $m$. The values of $b$ and $c$ are fixed in terms of $a$ and $m$ by the definition $m=b-ac$ and the normalization of $P(\zeta)$; we discuss the latter now. We use the normalization that for the initial distribution of couplings, $\int P(\zeta)d\zeta=1$, in contrast to previous work which defined $\int P(\zeta)d\zeta=3$, corresponding to the number of bonds per spin in the originally triangular lattice \cite{laumann2012strong}. In our case, $N\int P(\zeta)d\zeta$ is equal to $1/3$ times the number of bonds in the lattice at all RG times, so that $\int P(\zeta) d\zeta \leq 1$ for a planar graph (see Section~\ref{sec:geom_observables}). For the analytical work in Section~\ref{sec:analytical} we will use $\int P(\zeta) d\zeta = 1$ at all RG times.

\begin{figure}
    \centering
    \includegraphics[width=0.55\textwidth]{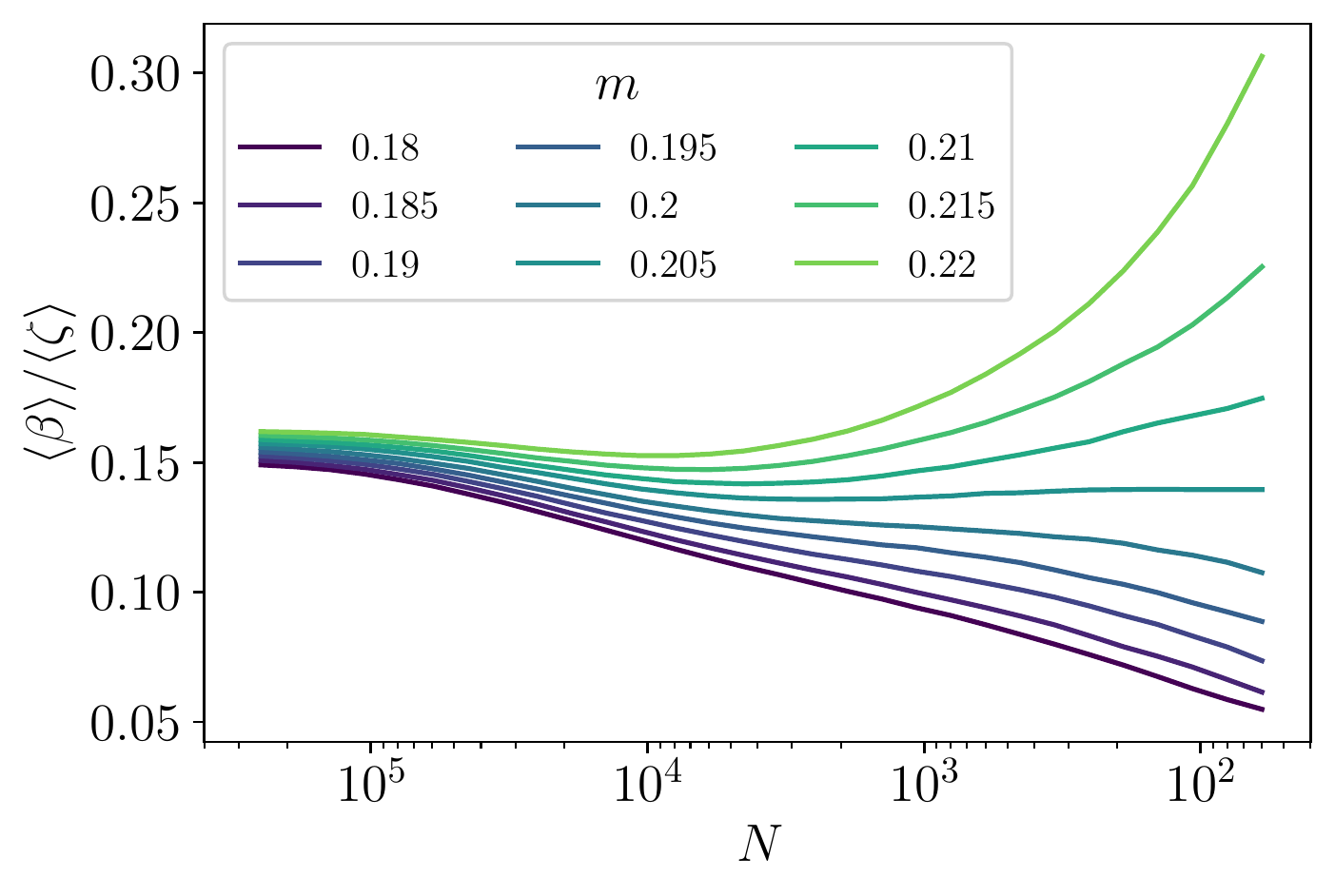}
    \caption{Numerical sweep to find the critical point which flows to infinite randomness. The ratio $\langle\beta\rangle/\langle\zeta\rangle$ flows to $0$ in the paramagnet and $\infty$ in the ferromagnet. At criticality it flows to some nonzero constant. For initial $P(\zeta) = \frac13(a+b\zeta)e^{-c\zeta}$ and $R(\beta) = e^{-\beta}$, we fix $a=0.1$ and vary $m = b-ac$. Criticality is achieved for $m=0.205$. Data are averaged over $100$ disorder realizations on a $500\times 500$ triangular lattice.}
    \label{fig:critical_separatrix}
\end{figure}

\begin{figure}
	\centering
    \includegraphics[width=0.9\textwidth]{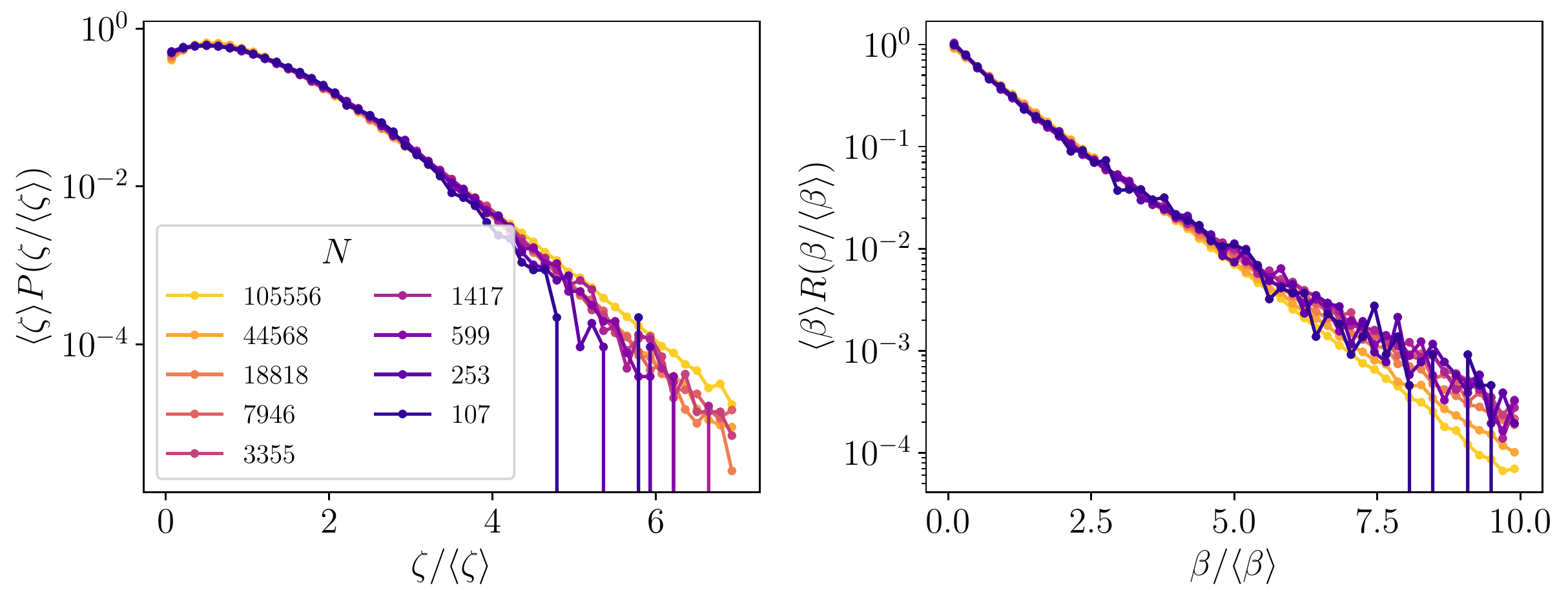}
	\caption{Scaling forms at criticality ($a=0.1$ and $m=0.205$) for distributions of couplings along the flow to infinite randomness, visualized by plotting the distributions of $\zeta$ and $\beta$ respectively normalized by their mean values. These reach a constant shape in the scaling regime of the RG.
	}  
	\label{fig:distributions}
\end{figure}

Fixing $a=0.1$ and varying $m$, we see in Fig.~\ref{fig:critical_separatrix} that there is a transition from $\langle\beta\rangle$ growing slower than $\langle\zeta\rangle$ to growing faster, as $m$ is increased. Infinite randomness is reached at the point where both scale the same way, here roughly $m=0.205$.
The fixed-point scaling forms of these distributions are seen in Fig.~\ref{fig:distributions}.
Since the SDRG is only asymptotically controlled, and indeed since different SDRG schemes lead to the same fixed point, the initial distributions that lead to the IRFP are not universal objects. Different RG schemes will have different initial transients which determine what initial conditions lead to infinite randomness.

\begin{figure}
	\centering
    \includegraphics[width=0.9\textwidth]{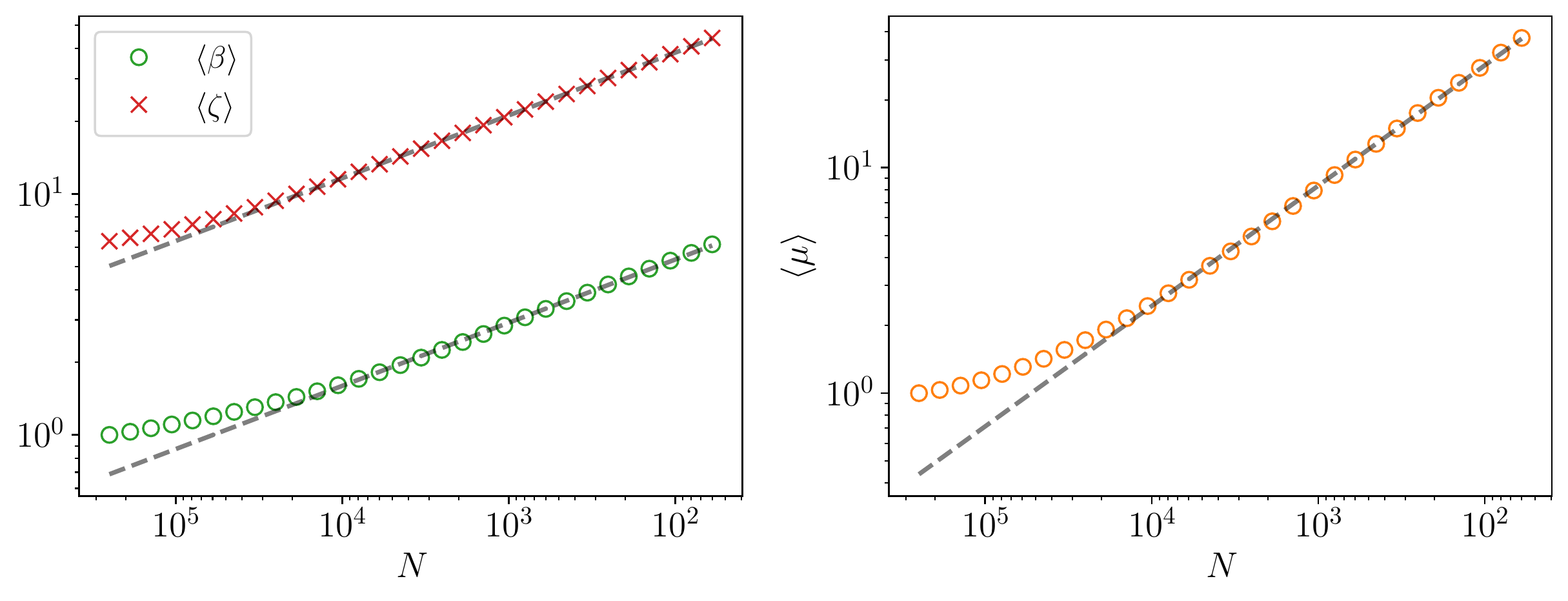}
	\caption{Numerical extraction of critical exponents characterizing the IRFP. The exponents $\psi$ and $d_f$ are defined by $\langle\beta\rangle,\langle\zeta\rangle\sim N^{-\psi/d}$ and $\langle \mu\rangle \sim N^{-d_f/d}$, with $d = 2$. Fitting power laws to our numerics after an initial transient (at the presumed critical point) gives values of $\psi=0.52\pm 0.04$, $d_f=1.06\pm0.03$.
	}  
	\label{fig:crit_exp}
\end{figure}

To further check the validity of the IRFP found by our scheme, we measure the critical exponents $\psi$ and $d_f$, which describe respectively the scaling of log energies with length scale and the fractal dimension of the ferromagnetic clusters (Fig.~\ref{fig:crit_exp}). There are numerous ways to fit the exponent $\psi$ from numerical data, since any measure of the width of $P(\zeta)$ and $R(\beta)$ should broaden as a power of $N$ in the same way at criticality.
Using the scaling of $\langle\zeta\rangle$ and $\langle\beta\rangle$ with $N$, we find $\psi=0.52\pm 0.04$, while the scaling of $\langle\mu\rangle$ with $N$ gives $d_f=1.06\pm0.03$ (Fig.~\ref{fig:crit_exp}). Alternatively we can fit an exponential form to $R(\beta)$ and extract the width this way (not illustrated), in which case we obtain $\psi=0.44\pm 0.05$, which is a less precise estimate for $\psi$ due to dependence on the range over which the fit is performed. In both of these cases the error bars refer to the variation in the critical exponents arising from different realizations of disorder, drawn from the distributions at the supposed critical point. These results for critical exponents are in agreement (within uncertainties) with previous work \cite{motrunich2000infinite, KovacsIgloi, laumann2012strong, Pich, KangSuperuniversality}.

\subsection{Geometrical observables}\label{sec:geom_observables}
\begin{figure}
    \centering
    \includegraphics[width=\textwidth]{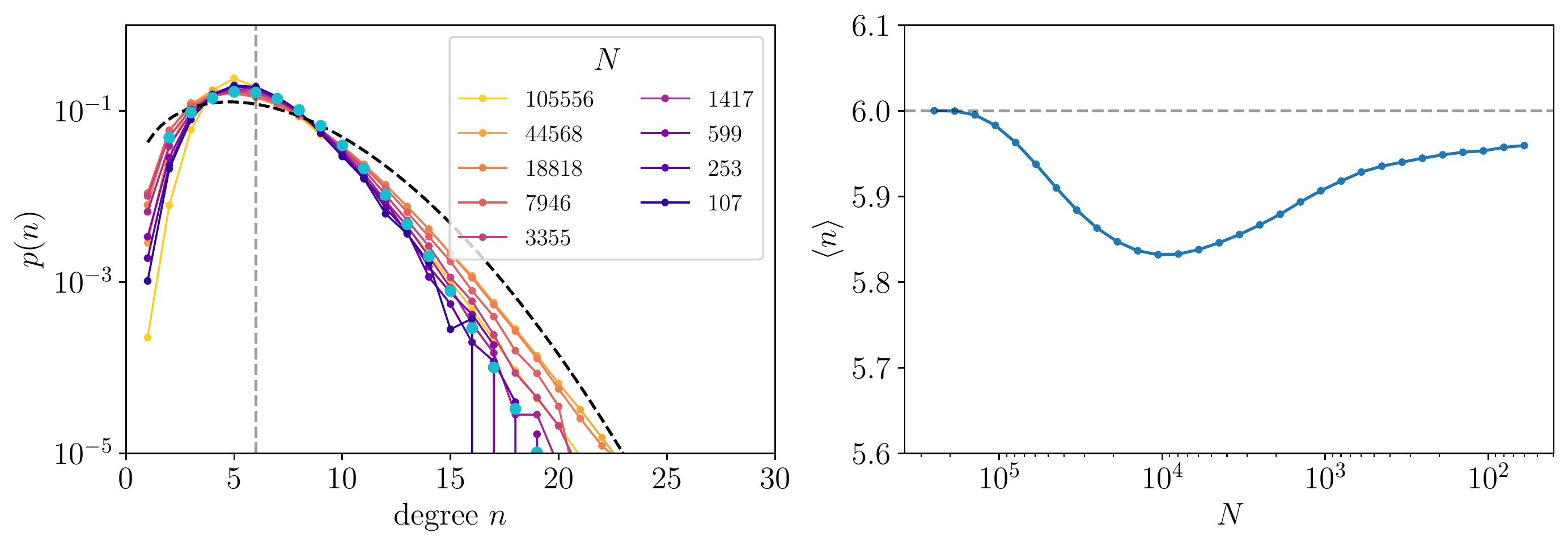}
    \caption{Distribution of spin degrees and mean degree in the interaction graph for various RG times, at criticality. On the left the distribution $p(n)$ is seen to reach a steady state. The vertical dashed line is at $n=6$. The light blue dots show a Poisson distribution, with slight modifications defined in Appendix~\ref{app:couplingRG}, and the dashed black line shows the modulated Gaussian envelope from Eq.~\eqref{eq:gaussian_pn}, plotted for $n\geq 1$. On the right is the mean degree of the graph over RG time, which goes below $6$ due to dangler production, but increases again (thus allowing us to neglect danglers) at late RG times.
    }
    \label{fig:degree_dist}
\end{figure}
Having established that our triangular RG scheme indeed leads to the IRFP of the (2+1)d TFIM, we can proceed to extract geometrical information about the renormalized lattice that is illuminated by the parsimony of our scheme.

The most basic feature of the graph is the distribution of degrees $p(n)$. Of particular interest is how this distribution evolves as the lattice is renormalized. 
In Fig.~\ref{fig:degree_dist} we see that, starting from a $500\times500$ triangular lattice with an initial $p(n) = \delta_{n6}$ (all vertices of a triangular lattice have six neighbors), $p(n)$ broadens to near a steady-state distribution, after decimating only two-thirds of the spins. Crucially the steady state is stable over orders of magnitude in $N$. This feature of the dynamics of $p(n)$ will be of critical importance in our understanding of the interplay between couplings and geometry to be discussed later.

The mean degree of the vertices for a triangulation (of a torus or an infinite plane) is $6$. This follows from the Euler characteristic $F - E + V = 0$  for the torus, the fact that the mean degree is given by $\langle n \rangle = 2 E / V$, and the fact that each vertex is shared by two faces on average in the triangulation, $F/V = 2$.  The latter observation follows because the sum of interior angles in a triangle is $\pi$ which accounts for half of a vertex. Combining these yields $\langle n \rangle = 6$. For a finite graph on a plane there will be corrections to this answer, but they approach zero in the limit of big $V$, i.e. big $N$. 

Notice that a finite toroidal graph with mean degree $\langle n \rangle = 6$ must be a triangulation since any other toroidal graph will have faces which are larger polygons and thus may be triangulated by the insertion of additional bonds. Therefore any non-triangulating toroidal graph has smaller mean degree as we can add edges without adding vertices to reach a triangulation. This demonstrates that any deviation from a mean degree of $6$ is equivalent to a fluctuation away from triangulations.

\begin{figure}
    \centering
    \includegraphics[width=0.5\textwidth]{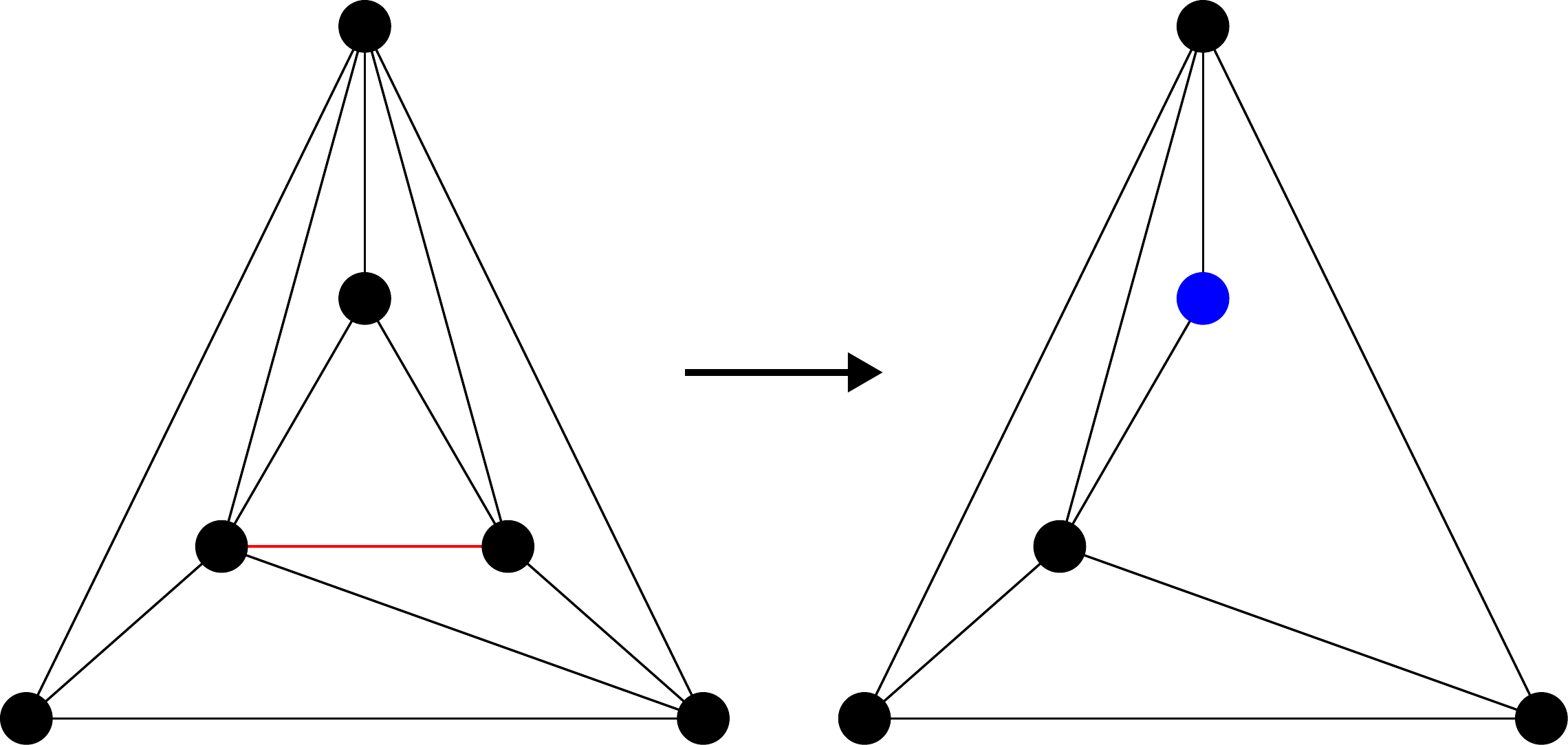}
    \caption{An illustration of a decimation that produces a ``dangler''. Our decimation scheme is not guaranteed to preserve triangulations of the plane. In the event that a bond is contracted for which the endpoints share more than two spins in common, a dangling spin (in blue here) --- more generally a dangling group of spins --- will be generated. This is tantamount to a face with size bigger than $3$ (here 4) which makes the mean degree of the graph drop below $6$. }
    \label{fig:dangler_production}
\end{figure}

One subtle feature of the triangular RG rule is that it does not perfectly preserve triangulations of the plane (or torus). As shown in Fig.~\ref{fig:dangler_production} we can perform a decimation starting from a triangulation, which leads to a non-triangulation. These non-triangulations are universally associated with spins which we call {\it danglers}. Danglers are produced whenever we contract an edge (via either site or bond decimation) whose two endpoints have more than two spins as common neighbors. A dangler-producing decimation makes the mean degree of the graph drop below $6$.

In the triangular RG the production of these danglers is very rare, as we have measured by looking at the mean degree of spins in the lattice, which never drops below $5.8$  (Fig.~\ref{fig:degree_dist}). We also see $p(n)$ is small but nonzero at $n = 1$ and $2$, which must come from danglers, as these degrees do not exist in a triangulation of a torus. Therefore, while the triangular RG rule does not perfectly maintain triangulations, as a practical matter the violation is very small. This property provides an important window into the correlations inherent in the RG, and can be understood via its tendency to decimate spins of lower degree, which we discuss in Section~\ref{sec:analytical} and Appendix~\ref{app:correlations}. 
In addition, starting from a square lattice with mean degree $4$, one observes a steady increase in the mean degree toward $6$, which is the largest possible mean degree for planar maps and is achieved only for triangulations.

Let us now discuss the justification of our choice of RG truncation. Ultimately the triangular RG is justified because it leads to the correct IRFP, to wit an IRFP with universal data that is in agreement with previous work. More qualitatively, the geometry and behavior of the couplings conspire to keep the RG well controlled, as we discuss here.

The proximity of the mean degree to 6, as shown in Fig.~\ref{fig:degree_dist}, provides support for the notion that characterizing our fixed-point geometry as random triangulations is a viable approximation.
Additionally, the distribution of degrees does not broaden significantly despite the possibility that, for instance, the variance could be unbounded: the tail at large $n$ decays faster than exponentially. These features imply that the geometric part of the triangular RG is consistent, so we may use it as a stable foundation to understand the coupling part of the triangular RG. 

One way to measure the error of approximation in ignoring the non-planar couplings is to look at how small the bonds we neglect are. When we decimate a spin $h$ of degree $n$ we take bonds $J_1 > J_2 > \dots > J_n$ and replace them with bonds $J_1 J_2 / h, J_1J_3 / h, \dots J_1J_n / h$ and so the strongest neglected bond is $J_2J_3/h$. If this bond is weaker than the bonds we do include, this lends some credence to the approximation.

This motivates the following definition for the average number of bonds which can be added before we neglect a bond stronger than one we are adding,
\begin{equation}
    t_{\text{cutoff}}(n) = \mathbb{E} \left[\max\{i ~|~ J_1J_i \geq J_2J_3, ~ i \leq n\} \right]
\end{equation}
where $n$ is the degree of the spin under consideration, and $J_i$ are the ordered bonds attached to that spin which are drawn independently at random from the distribution over couplings. Despite the fact that the couplings are in general correlated this quantity will give a sense for why the triangular RG is sufficient instead of a more complicated scheme. We find, for $\zeta$ distributed according to either an exponential or the numerically observed fixed-point $P(\zeta)$, that $t_{\text{cutoff}}(6) \approx 4.5$, with a very slow upward trend as $n$ increases above $6$. Considering that decimating a spin of degree 6 usually requires us to generate only 3 new bonds, we see that typically we do not add the bonds out of order. As the distribution of degrees sharply decays for $n > 6$, we rarely decimate spins of very large degree, which is the setting in which the triangular RG could make large errors, namely when $t_{\text{cutoff}}$ becomes small compared to the number of bonds generated.

\subsection{Two-stage nature of the triangular RG}

\begin{figure}
    \centering
    \includegraphics[width=0.5\textwidth]{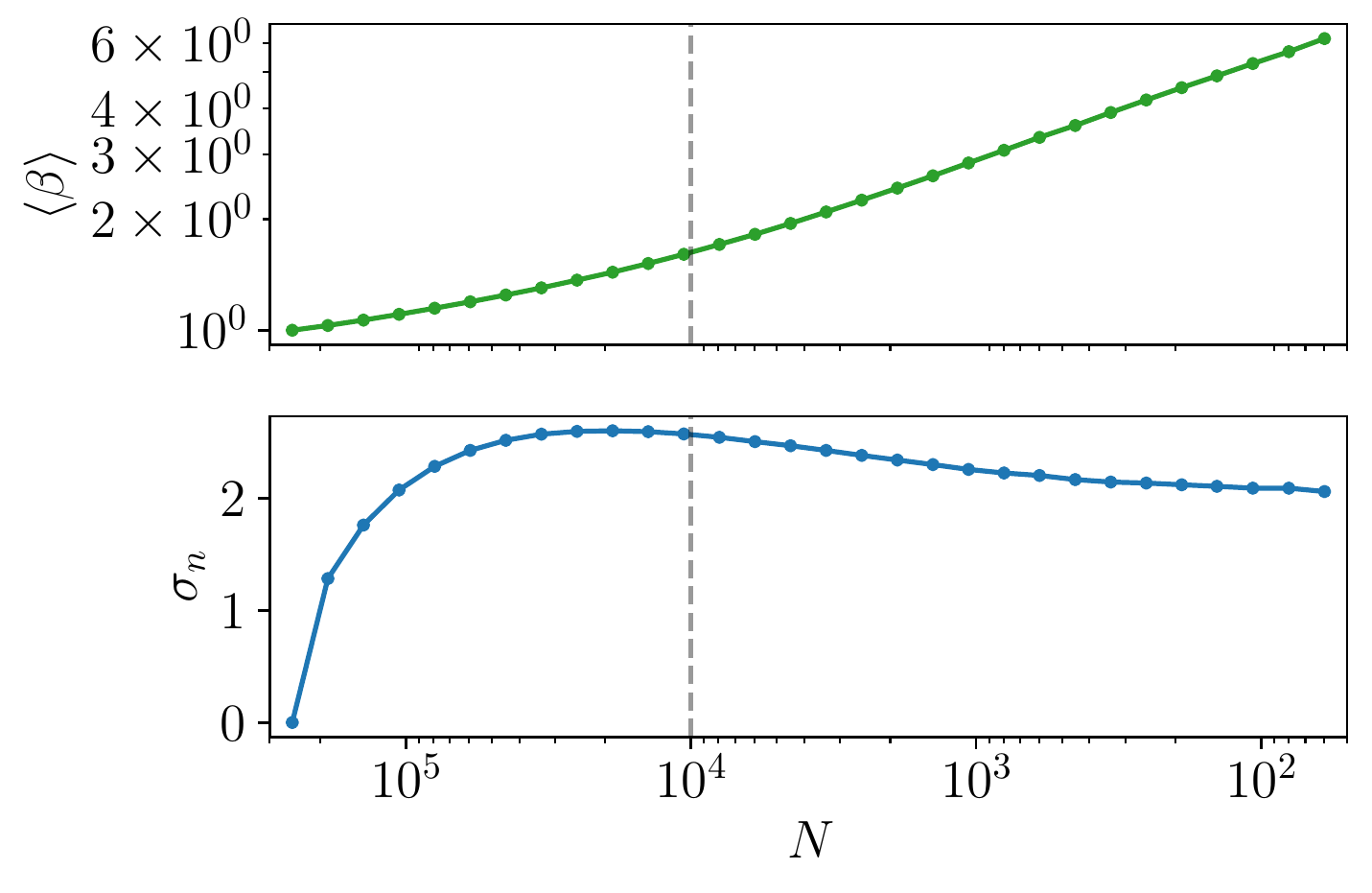}
    \caption{Numerical observations summarizing the two-stage nature of the triangular RG. The first stage mainly consists of the renormalization of the lattice, seen through the broadening of $p(n)$ from the initial $p(n)=\delta_{n6}$ for a triangular lattice, and concomitant increase in $\sigma_n$, the standard deviation of $p(n)$. After the dashed vertical line, with $\sigma_n$ close to constant, the RG approaches the infinite-randomness scaling regime, with $\langle\beta\rangle$ scaling as $N^{-\psi/d}$.}
    \label{fig:two_stage_rg}
\end{figure}

The results presented thus far constitute the majority of our numerics in the triangular RG; some further comments on correlations, particularly those between degrees and couplings, may be found in Appendix~\ref{app:correlations}, which also contains simulations of the flow of geometry under random decimations (divorced from the Ising couplings which have determined the flow of geometry here). We now discuss the structure of the triangular RG which we will substantiate with analytical toy models in the next section.

Fig.~\ref{fig:two_stage_rg} summarizes the most pertinent results. The two panels plot the width of the log field distribution $R(\beta)$, measured by $\langle \beta\rangle$, and the standard deviation of the degree distribution $p(n)$, which we call $\sigma_n$, versus  $N$ through the RG. We interpret the RG as being composed of two stages, the crossover between which is delineated in Fig.~\ref{fig:two_stage_rg} by the dotted line at $N = 10^4$. This crossover is the point at which the geometrical distribution (of which $p(n)$ is the most important representative) has saturated and does not change for later RG times, i.e. smaller values of $N$.

We start with a distribution $R(\beta) = e^{-\beta}$, so that $\langle \beta \rangle = 1$, and a triangular lattice which has $p(n) = \delta_{n6}$ and thus $\sigma_n = 0$. As the RG goes on, both distributions broaden, but approach long-time limits of somewhat different characters. After some RG time, $\sigma_n$ asymptotes to a constant. We have already commented on the evolution of $p(n)$: we know that it saturates to a distribution with width of order $1$ and quickly decaying   at large $n$, and indeed we used these facts to rationalize the effectiveness of the triangular RG. The $\beta$ distribution broadens to infinite randomness, and enters its scaling regime only after $\sigma_n$ has stabilized.

The first stage of the RG, therefore, may be seen as a purely geometrical process in which $p(n)$ approaches its steady-state form. During this stage of the RG the evolution of the coupling distributions may as well be ignored, although there is a caveat due to geometry-coupling correlations: as we will show in Section~\ref{sec:analytical}, if there were no correlations between the geometry and the couplings, a steady-state $p(n)$ of the form we observe would not result. We will show this by comparing toy models with and without these correlations present; also see Appendix~\ref{app:correlations}.

In the second stage of the RG, while the graph of course changes, we conjecture that it belongs to an invariant \emph{ensemble} of graphs, represented by the steady-state $p(n)$ (and in principle more complicated graph data which we will ignore for simplicity), and flows ``ergodically'' through it. Now the couplings flow on top of this fixed-point graph ensemble, and quickly approach infinite-randomness scaling. It is natural that infinite-randomness scaling for couplings appears only after the geometry has saturated: it is the geometry which governs the frequencies of the various decimation processes that broaden the coupling distributions. Also, coupling-geometry correlations have evolved and saturated so that the steady-state $p(n)$ can be maintained. 

Finally, note in the second panel of Fig.~\ref{fig:degree_dist} that in the second stage of the RG the mean degree, which had at early RG times dropped to $\langle n \rangle \approx 5.8$ due to dangler production, increases towards $6$ again. This is rather striking since it signals a tendency of the scaling regime (with all correlations present) to correct for non-triangulations in a self-consistent manner, and provides further evidence that thinking of the fixed-point geometry of this RG as triangulations is accurate.

\section{Analytical models}\label{sec:analytical}

Having outlined our understanding of the two stages of the triangular RG as seen in the numerics, we now discuss analytical models for both stages, which result in fixed points for the degree distribution and the coupling distributions respectively. These agree with the numerical observations qualitatively and, for analytically computed critical exponents, approximately quantitatively.

\subsection{Geometrical steady state}

We discuss the mechanism by which the distribution of degrees $p(n)$ changes as a function of RG time, and a toy model including coupling-geometry correlations which reproduces the correct form of the steady-state $p(n)$ observed in the numerics.

Whenever a decimation (either site or bond decimation --- as we have seen, their actions on the graph are identical) takes a triangulation to a triangulation, the changes that are effected on the list of degrees are as follows: the spins on either side of the edge that is contracted are joined up to create a new spin of bigger degree: if the two degrees were $n_1$ and $n_2$, the new degree is $n_1 + n_2 - 4$. The two spins that were both neighbors with the two spins thus joined, say with degrees $n_3$ and $n_4$, are now each joined to one new spin, and so their degrees go to $n_3 -1$ and $n_4 - 1$. If the mean degree before the decimation was $(\sum_I^{N-4} n_I + n_1 + n_2 + n_3 + n_4)/N = 6$, then the new mean degree is  $(\sum_I^{N-4} n_I + n_1 + n_2 - 4 + n_3 - 1 + n_4 - 1)/(N -1) $, which is also $6$, as required by the fact that graph remains a planar triangulation. This exercise also shows exactly how danglers are produced: if the edge being contracted had more than two common neighbors, the mean degree would drop below $6$. Therefore, since the graph is still planar, it must have at least one face of size bigger than $3$.

Let us decompose this into two processes. The first is a shift in the denominator by $-1$ without changing the numerator, which may be seen as the bunching up of spins with no consideration of bonds. This produces a change in the mean degree of  $\delta \langle n \rangle \approx  \langle n \rangle /N$. The second is a shift in the numerator, which models the removal of double-counted bonds without changing the number of spins; this changes the mean by $\delta \langle n \rangle \approx -6/N$. These contributions add up to zero if and only if $\langle n \rangle =  6$.  

Now we generalize the coefficients of these equations (allowing $n$ to be continuous) to
\begin{equation}
    \delta \langle n \rangle = \alpha_+ \langle n \rangle - \alpha_-.
\end{equation}
We can consider similar evolution equations for general moments $\langle n^m \rangle, m =1, 2,3, \dots$: the easiest way to control this calculation is through a further simplification: subtract $\alpha_-/N$ off the degrees of all $N$ spins in the numerator, rather than ${\cal O}(1)$ amounts off an ${\cal O}(1)$ number of spins. 

These manipulations lead us to
\begin{equation}
    \delta \langle n^m \rangle = \alpha_+ \langle n^m \rangle- \alpha_- m \langle n^{m-1} \rangle .
\end{equation}
These equations have built in a further approximation --- which is incorrect and will need to be amended --- namely that all spins are equally likely to be involved in a decimation; in other words they are equally likely to be involved in the numerator subtraction which yields the term proportional to $\alpha_-$.

To find the steady state, we impose that all moments are unchanged by a decimation, $  \delta \langle n^m \rangle = 0$. We find $\langle n \rangle = \alpha_-/\alpha_+ $, which we set equal to $6$, and for $m\geq 2 $, $\langle n^m \rangle =6 m   \langle n^{m-1} \rangle = 6^m m! $. The distribution that has these moments is the exponential, $p(n) = \frac{1}{6} e^{-n/6}$.

This explains why the correlationless geometrical RG numerics of Appendix \ref{app:correlations} --- where decimations are performed randomly on a given graph with no couplings present to provide a measure as a function of degree on these decimations --- sees an exponential $p(n)$. The comparison is subtle since that RG also produces a lot of danglers. It is easy to see, however, that the abundance of danglers is linked with the abundance of small degrees in the triangulation, which in turn is linked with $p(n)$ being monotonically decreasing.

Since we understand which correlations generate the difference between the answer above and the numerically observed one, we can build this physics into our model. As we have mentioned above and numerically verify in Appendix \ref{app:correlations}, spins with small degrees are more likely to be involved in decimations since they typically host larger $h$'s and are adjacent to larger $J$'s. Let us therefore carry out degree subtractions at degree $n$ with weights $w_n$. The spin-joining events that scale moments by $\alpha_+$ are agnostic to which spins were involved, since they only change the denominator. Thus we write
\begin{equation}
    \delta \langle n^m \rangle = \alpha_+ \langle n^m \rangle  - \alpha_- m \langle n^{m-1}w_n \rangle .
\end{equation}
We solved the version of this equation with all $w_n = 1$, but would now like $w_n$ to decrease with increasing $n$. The correct choice of $w_n$ can in principle be obtained from measurements in the fully correlated RG; here we consider the simplest solvable choice $w_n = 1/n$. We have for the steady state $\langle n^m \rangle  = m (\alpha_-/\alpha_+)  \langle n^{m-2} \rangle$. Fixing $\langle n\rangle = 6$, we see that the recursion is satisfied by 
\begin{equation}
\label{eq:gaussian_pn}
    p(n) = \frac{1}{\alpha_-/\alpha_+} n \exp\left(-\frac{n^2}{2 \alpha_-/\alpha_+}\right),
\end{equation}
where $\alpha_-/\alpha_+ = 72/\pi$.

This distribution has a hump, and its decay at large $n$ is stronger than exponential. Both of these features agree with the $p(n)$ observed in the numerical RG (with all correlations present), as can be seen in Fig.~\ref{fig:degree_dist}. It also agrees with the geometrical RG of Appendix~\ref{app:correlations} where spin decimations are repeatedly attempted on a graph but a given decimation is ``accepted'' with probability $1/n$.  

The weight $w_n = 1/n$ is not meant to be an accurate representation of the actual measure on decimations in the RG that retains all correlations; it is simply a solvable example that reproduces the most salient features of the numerically observed $p(n)$. Tweaking $w_n$ one can of course get better quantitative agreement.

Having understood the effect of geometry-coupling correlations on the geometry, we may fix the geometric background and study the infinite-randomness flow of coupling distributions of Hamiltonians living on this steady-state geometry.

\subsection{Correlationless flow of couplings}

Once the geometry --- for which the degree distribution $p(n)$ is a coarse proxy --- has reached its steady state, the distributions of log couplings $P(\zeta)$ and $R(\beta)$ continue to broaden, and indeed approach infinite-randomness scaling, as defined in Section~\ref{sec:triangular}. We shall now describe a toy model in which this broadening of distributions, and their approximate scaling forms, can be seen explicitly.

The model takes as input a steady-state degree distribution $p(n)$, and describes the broadening of $P(\zeta)$ and $R(\beta)$ under the triangulation RG rules through RG equations similar to Fisher's~\cite{fisher1995critical}, but with added complexity due to the removal and production of many bonds when a spin is decimated.

The most crucial approximation here is the truncation of all correlations among the couplings and most correlations between the couplings and the geometry. For example, upon a decimation that produces bonds with strengths $\zeta_1 + \zeta_2$ and $\zeta_1 + \zeta_3$, which are correlated, our equations will in fact make new independently distributed bonds corresponding to random variables with the same marginal distributions as $\zeta_1 + \zeta_2$ and $\zeta_1 + \zeta_3$. We therefore also never have to think about the possibility of strong transverse fields having strong bonds next to them, and other correlations which in principle exist at the fixed point (see Appendix~\ref{app:correlations} for further discussion). As we have discussed above, coupling-geometry correlations, such as the fact that spins with smaller degrees carry larger transverse fields, are known to be essential in order to maintain the steady state $p(n)$ in the first place. These correlations are implicitly taken into account by not changing $p(n)$ during the flow. However, we do not actually compute a flow of couplings including correlations with $n$. With these truncations, the only aspect of the geometry that appears in the RG equations is in fact $p(n)$, as this governs the number of bonds destroyed/created when spin decimations take place. 

The other simplification we have made concerns cases in which newly produced bonds already exist in the graph. The numerical RG step uses the bigger of the new and old bonds and entirely discards the smaller one (Fig.~\ref{fig:triangrules}). However, our RG equations will model a process in which the old bond is always replaced by the new one, regardless of whether the new bond is stronger or weaker. We have no way to justify this approximation other than that it makes the system more analytically tractable.

With all of these caveats in place, the RG equations are
\begin{widetext}
\begin{equation}\label{eq:secPequation}
    \frac{\partial P }{\partial \Gamma} = \frac{\partial P}{\partial \zeta} + \frac{R_0}{3} \sum_n p(n) \left[  -(n+2) P  + \sum_{i = 2}^n \int_0^\zeta d\zeta_1 P^{(n)}_{1i}(\zeta_1, \zeta - \zeta_1) \right]  + (P_0 +  R_0 ) P,
\end{equation}
\begin{equation}\label{eq:secRequation}
  \frac{\partial R}{\partial \Gamma} = \frac{\partial R}{\partial \beta}  + 3P_0 \int_0^\beta d\beta_1 \, R(\beta_1) R(\beta - \beta_1) + (-3P_0 + R_0)R.
\end{equation}
\end{widetext}
The factors of $3$ come from the fact that there are three bonds per spin in a triangulation, i.e. the average degree is $6$, and that we use the normalization $\int d\zeta P(\zeta) = 1$. We provide a careful derivation of the above equations in Appendix \ref{app:couplingRG}.

Eq.~\eqref{eq:secRequation} is identical in form to its one-dimensional analogue, \eqref{eq:1dRGR}, reflecting the identical ways in which the distribution of transverse fields is modified; compare Figs.~\ref{fig:1drules} and \ref{fig:triangrules}. Eq.~\eqref{eq:secPequation}, on the other hand, differs from \eqref{eq:1dRGP}. The term that destroys and creates bonds during a spin decimation has the following structure: first, the spin will have degree $n$ with probability $p(n)$, so we must sum over all $n$ with weights $p(n)$. If the decimated spin has degree $n$, we destroy $n+2$ bonds: the $n$ bonds connected to the spin, and also two that are the sides of the polygon connected to $J_1$ ($J_L$ and $J_R$ in Fig.~\ref{fig:triangrules}). Then we create $n-1$ bonds with log strengths equal to $\zeta_1 + \zeta_i$, $i \in \{2,3,\dots, n \} $. The joint distributions of $\zeta_1$ and $\zeta_i$ (the so-called order statistics of $P$) are $P^{(n)}_{1i}(\zeta_1, \zeta_i) $, and their explicit form is given in Appendix~\ref{app:couplingRG}.

With a certain $p(n)$ (similar to Poisson) that approximates the numerically observed degree distribution (see Fig.~\ref{fig:degree_dist}), progress on solving these equations can be made analytically. Indeed we employ a scaling ansatz with variational parameters whose optimal values yield distributions that agree with numerical observations. For more details, see Appendix~\ref{app:couplingRG}.

The upshot is that this simple model contains an IRFP (and flows of the paramagnetic and ferromagnetic sorts on the two sides --- these are not discussed here) with critical exponents $\psi \approx 0.6$ and $d_f \approx 1.3$, relatively close to the critical exponents observed in the numerics. We should not make too much of this numerical agreement, however: as we relax the various violent approximations that went into the model, the numerical estimate for the critical exponents will get better, but there is no reason to believe that the approach to the correct critical exponents will be monotonic. We do not know how to make analytical progress beyond this simplest correlationless model.

\section{Discussion}\label{sec:discussion}

Using a truncation of RG rules that simplifies the geometry of the problem, we have studied the structure of the infinite-randomness fixed point of the critical (2+1)-dimensional transverse-field Ising model. We now discuss several open questions and speculate about possible answers.

First, we are naturally interested in an extension of our analytical results to more complicated models where the evolution of some geometry-coupling correlations can be kept track of explicitly with joint distributions such as $R(\beta, n)$ for log field $\beta$ on a spin of degree $n$. An even simpler extension of our work would at least keep track of a set of expectation values such as $\langle\beta_i n_i \rangle$: it would be desirable to find a set of quantities such that their evolutions can be solved for self-consistently, without the need for putting in correlations (such as non-trivial $w_n$) by hand. We have been unable to do this for the TFIM. On the other hand such an approach is unlikely to be able to find the full scaling forms of $P(\zeta)$ and $R(\beta)$.

It is natural to ask whether TFIM in higher spatial dimensions hosts infinite-randomness fixed points. The answer from numerical SDRG seems to be affirmative~\cite{motrunich2000infinite, KovacsIgloi3d}, and the critical exponents are apparently different from the two-dimensional case. We believe that a three-dimensional RG truncation, which would perhaps be best interpreted approximately as tetrahedra that ``tile'' three-dimensional space, is the appropriate way of attacking this problem. (In spatial dimensions bigger than 4, disorder is not perturbatively relevant at the clean critical point~\cite{Harris}, so while it is still possible, and numerically plausible~\cite{KovacsIgloi3d}, that an IRFP is hosted in large spatial dimensions, in this scenario the true RG flow would have a much more complicated form than in lower dimensions, and the one-parameter cartoon of the flow being characterized simply by the disorder strength could fail drastically.)

There are models for which the initial graphs are not lattices where the full (second-order perturbation theory) RG rule nevertheless generates a class of graphs that are closed under the RG. For example, site decimations on a tree will immediately generate loops, but the graph thus generated (and all graphs that can be generated by further site/bond decimations) will be composed of all-to-all clusters (``simplices'') arranged in a corner-sharing fashion such that if each simplex is regarded as a big vertex, there are no loops joining these big vertices up. These simplicial trees may well host IRFPs, and would be one way to think about possible IRFPs for the TFIM on a tree (for different approaches see e.g. Refs.~\cite{MonthusTreeBoundary, MonthusTreeMBL, MonthusGarelPolymer, DimitrovaMezard}). However, the geometrical problem here is richer than in our study of the lattice with triangular truncation, since in the absence of truncations the geometrical and coupling distributions grow infinitely broad together --- there is no two-stage simplification --- and geometry-coupling correlations are expected to be extremely important. We leave the study of these more complicated IRFPs for future work.

The connection of this work to experiment is tenuous, given that local moments in lightly doped semiconductors are not in an Ising limit, and certainly have no reason to be tuned near criticality. The more relevant problem will have Heisenberg exchange with big randomness. For a large range of energies SDRG reasoning yields excellent agreement with experiments, but it is expected that the eventual fixed point actually has a finite-randomness character (e.g. with conventional finite-$z$ scaling)~\cite{BhattLee, westerberg1997low, KimchiLee}. What geometrical principles might govern these finite-randomness fixed points is unknown: certainly truncations of the sort we have done can lead to spurious IRFPs as they drop bonds that would make the coupling distribution less broad than the truncated one. Besides, as we have stressed, SDRG is ultimately uncontrolled if the fixed point it reaches is not at infinite randomness, so it would be desirable to formulate RG schemes that are asymptotically exact with a finite disorder distribution.

Finally, while our perspective on the low-energy sector of gapless theories with quenched randomness is via the calculational power afforded by RG, one would eventually like to find kinematic principles that govern the fixed point, in analogy to the role conformal field theory plays at critical points in clean statistical mechanics. Recent work has aimed at starting to understand such a framework using tools from the physics of topological phases~\cite{MaWang, StrangeCorrelator}, but a general theory, including a connection to SDRG, remains an open issue.

\acknowledgments
We are grateful to D.S. Fisher, D.A. Huse, C.R. Laumann, O.I. Motrunich,  P.A. Nosov, N. O'Dea, and J.J. Yu   for discussions and correspondence. AP is supported by a Stanford Graduate Fellowship. AC thanks the Stanford Graduate Fellowship and Xiaoliang Qi's Simons Investigator Award (ID: 560571) for their support. AM is grateful for support from NSF grants PHY-1607606 and  PHY-2210386.
We thank the Stanford Sherlock cluster for computing resources.

\appendix

\section{Geometrical RG and correlations}\label{app:correlations}

\subsection{Correlationless geometrical RG}
One feature of the SDRG in two dimensions which is not present in one dimension is the development of correlations between couplings generated by second-order perturbation theory (e.g. multiple bonds can be generated involving the same decimated bond) and between couplings and geometry. However, we can imagine a simplified RG which essentially divorces the two processes of lattice renormalization and coupling renormalization, following our understanding of the actual RG as a two-step process. In this simplified RG we only study the lattice geometry, leaving the bonds and fields untracked.

Without the backing of fields and bond strengths, we supply our own rules on how to choose vertices/edges for decimation with the aim of understanding the behavior of the full SDRG. Note that in the triangular RG the operations of spin and bond decimation act identically with respect to graph geometry, as both operations correspond to the contraction of an edge. Therefore specifying a distribution over edges would specify a measure for decimations in this geometrical RG. We induce a measure over edges by the following procedure: pick a remaining vertex according to a specified measure on vertices (discussed below), and uniformly at random choose a bond next to it.

\begin{figure}
    \centering
    \includegraphics[width=.49\textwidth]{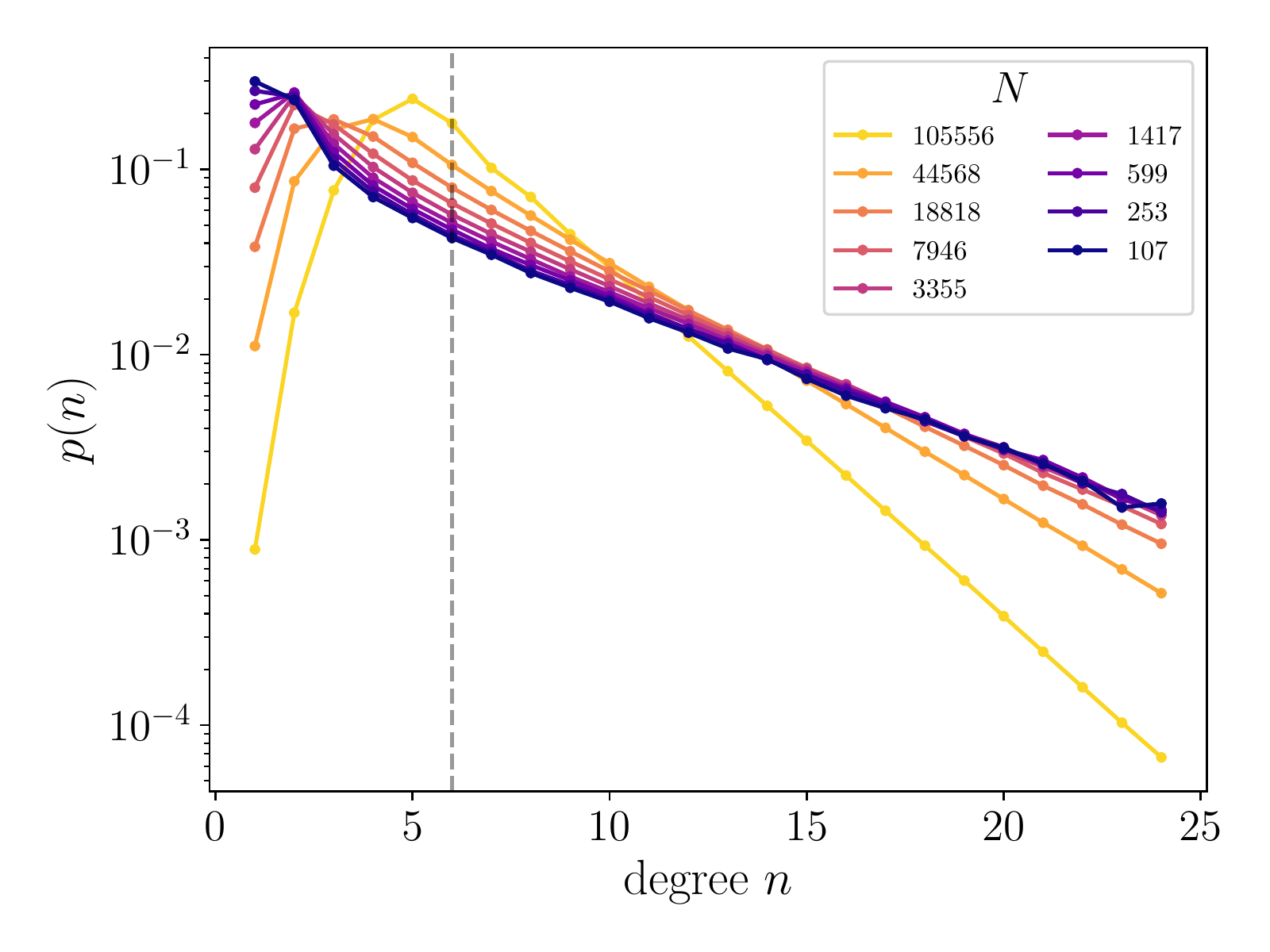}
    \includegraphics[width=.49\textwidth]{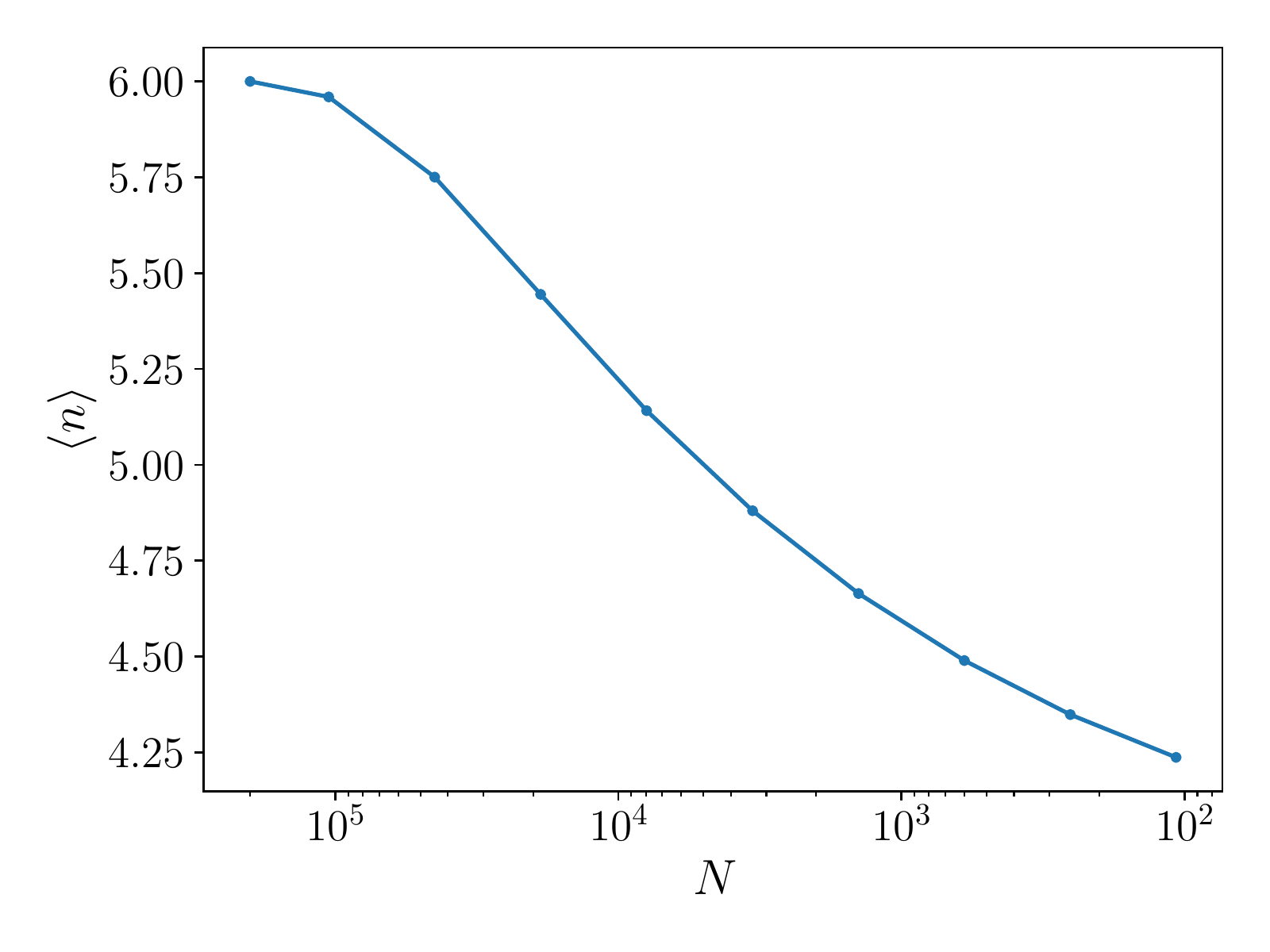}
    \caption{Degree distribution (left) and mean degree (right) for various RG times in a purely geometrical RG. The decimation rule repeatedly picks spins uniformly at random and contracts random edges next to them; couplings on the graph are immaterial. The initial graph is a $500\times500$ triangular lattice. Notice that the average degree drops over the course of the RG, and correspondingly the distribution of degrees approaches an exponential-like distribution.}
    \label{fig:degree_distributions_nocull}
\end{figure}

The simplest choice for the aforementioned measure on vertices is uniform. One might expect that this RG has a very different character than the fully correlated RG. Indeed we see that it is much more likely to produce danglers, and as a result the mean degree drops precipitously over the course of the RG, as we can see in Fig.~\ref{fig:degree_distributions_nocull}. Relatedly, the degree distribution $p(n)$ is a monotonically decreasing function of $n$, in contrast with the observed $p(n)$ in the RG with correlations. This big departure from the correct distribution signals that this procedure neglects important correlations; we now discuss how to build these into the geometrical RG.

\subsection{Correlations }
\label{sec:app_correlations}
\begin{figure}
    \centering
    \includegraphics[width=\linewidth]{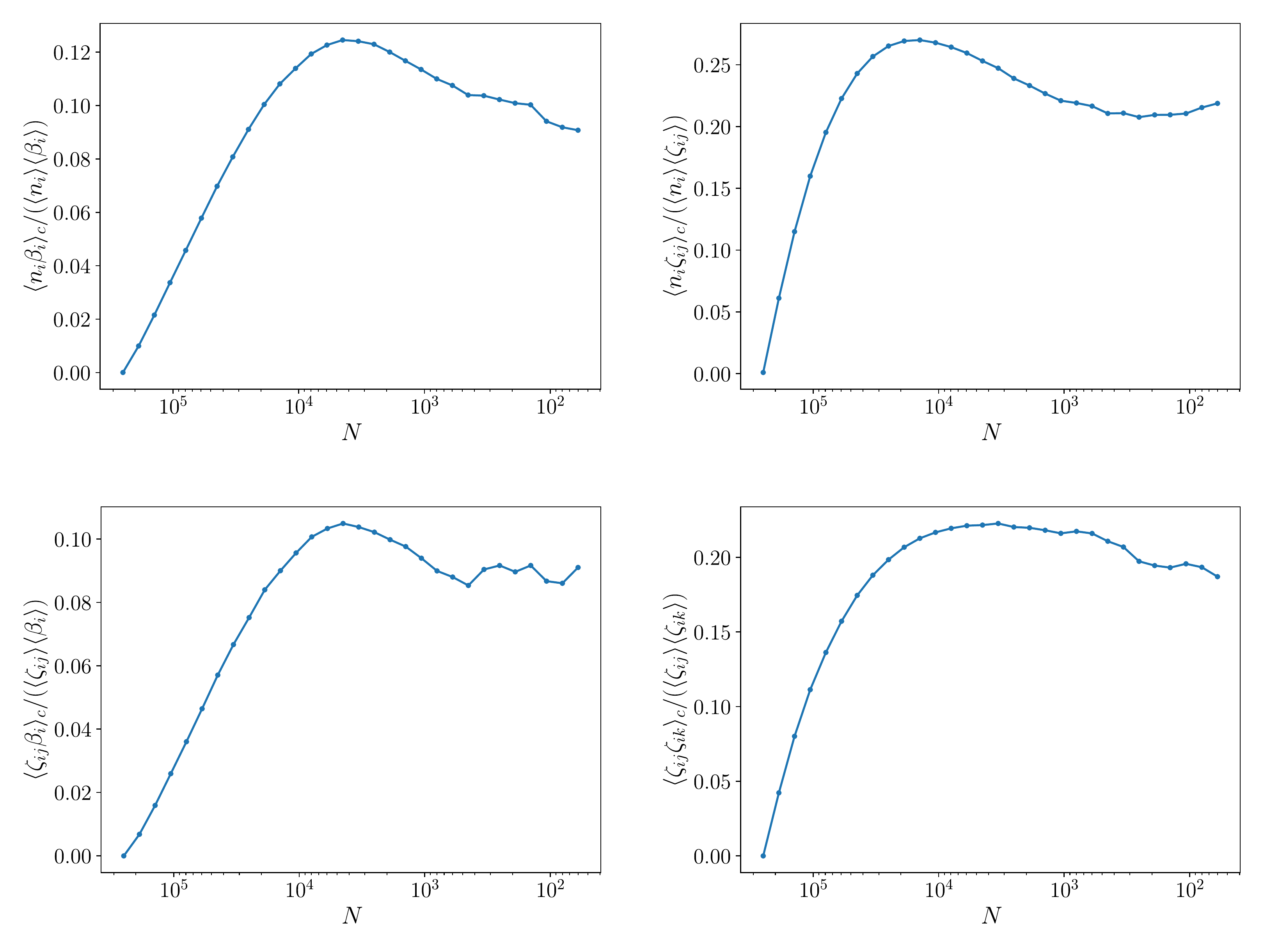}
    \caption{The development of correlations during the triangular RG. Each  of the connected correlations is normalized by dividing by the means of the random variables in the correlation. The top left shows the correlation between the degree of a spin and the log field present on that site, the top right shows the correlation between the degree of a spin and the value of a log bond attached to that site, the bottom left shows the correlation between a log field on a site and the log bonds attached to it, and the bottom right shows the correlation between two different ($j \neq k$) log bonds attached to the same site. As in our other numerics we start with a $500\times 500$ triangular lattice.}
    \label{fig:correlations}
\end{figure}

\begin{figure}
	\centering
    \includegraphics[width=0.9\textwidth]{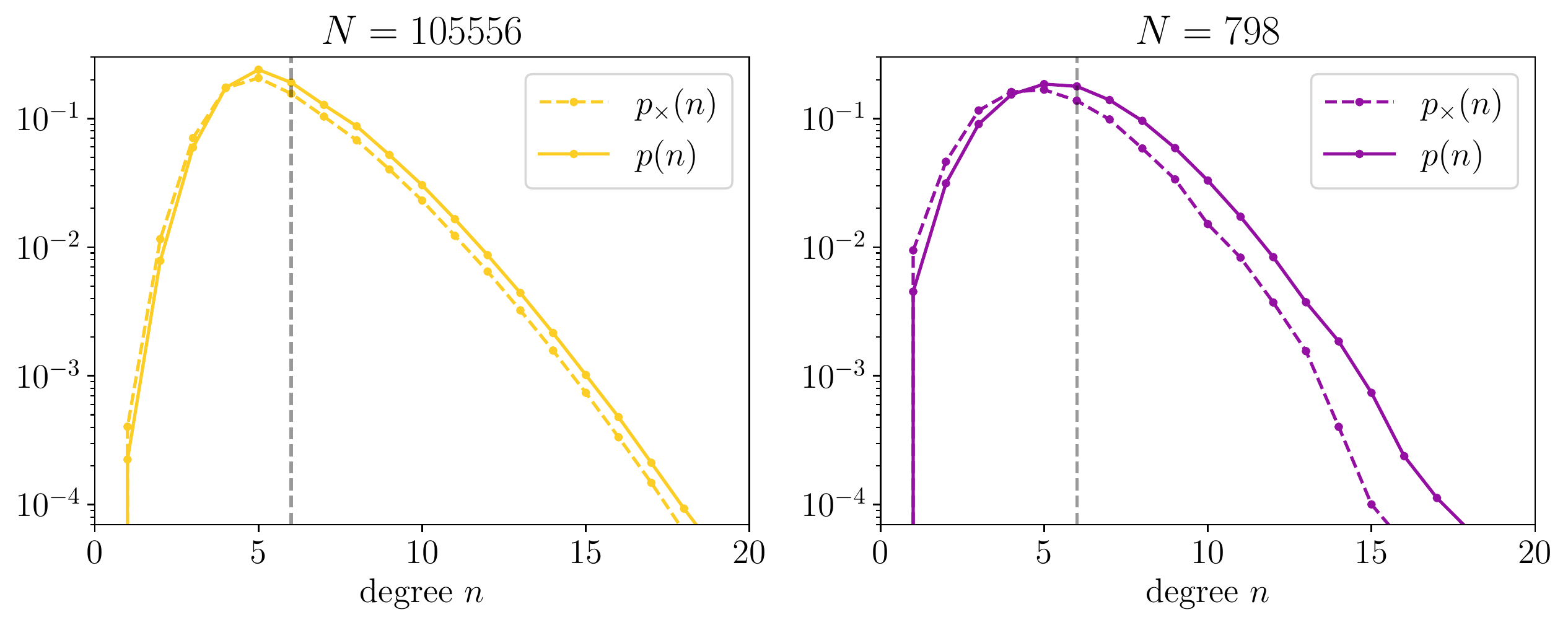}%
	\caption{Comparison at two different RG times between $p(n)$, the distribution of all degrees, and $p_\times(n)$, the distribution of degrees of spins decimated between the current $N$ and the next recorded $N$.  Though initially (when $N$ is large) spins are decimated with frequencies uncorrelated with their degrees, the correlations that build up between fields and geometry eventually cause low-degree spins to be decimated more than high-degree spins.
	}  
	\label{fig:decim_degree_dist}
\end{figure}

To understand how likely any vertex or edge is to be chosen we must understand how likely it is for the largest field or bond (in the actual triangular RG with couplings) to reside on it. The simplest local quantity these probabilities could depend on is the degree of the vertex, or of the vertices adjacent to that edge. The simplest type of correlation which would generate a non-uniform distribution of decimations over the graph would therefore be a correlation between the transverse field and degree of a spin, or between the strength of a bond and the degrees of the spins at the ends thereof.

We calculated the connected correlations $\langle \beta_i n_i \rangle_c/(\langle \beta_i \rangle \langle n_i \rangle)$, $\langle \zeta_{ij} n_i\rangle_c / (\langle \zeta_{ij} \rangle \langle n_i \rangle)$, $\langle \zeta_{ij} \beta_i\rangle_c/(\langle \zeta_{ij}\rangle\langle \beta_i \rangle)$, and $\langle \zeta_{ij}\zeta_{ik} \rangle_c / (\langle \zeta_{ij} \rangle\langle \zeta_{ik}\rangle)$ and found that they are positive and of order $1$. The dynamics of some important correlations over the RG flow are shown in Fig.~\ref{fig:correlations}. These imply that spins with larger degrees are more likely to have larger $\beta$'s on them and therefore smaller fields. Similarly bonds whose adjacent spins have large degrees are more likely to have small couplings on them. From a dynamical perspective this makes sense since decimation simultaneously increases the degree of some spins and also leads to smaller fields on them or smaller bonds attached to them. We can also see that these correlations are developed quickly as the geometry changes in stage 1 of the RG, and then remain relatively constant afterwards. This shows that the separation of the RG into two stages is self-consistent; in stage 2 of the RG these steady-state correlations maintain the steady-state geometry, as we have outlined in Section~\ref{sec:analytical}.

Another basic measure of correlations between geometry and couplings is to look at how the distribution of {\it decimated }  spin degrees $p_\times(n)$ differs from the overall $p(n)$. In Fig.~\ref{fig:decim_degree_dist} we see that although in the initial RG stages the two are similar (since correlations have had no time to develop), in the later stages they show a pronounced difference, with the distribution of decimated spins $p_\times(n)$ having a preference for spins with lower degree, indicating again that $n$ and $\beta$ for a spin are positively correlated.

\subsection{Correlated geometrical RG}

\begin{figure}
    \centering
    \includegraphics[width=.49\textwidth]{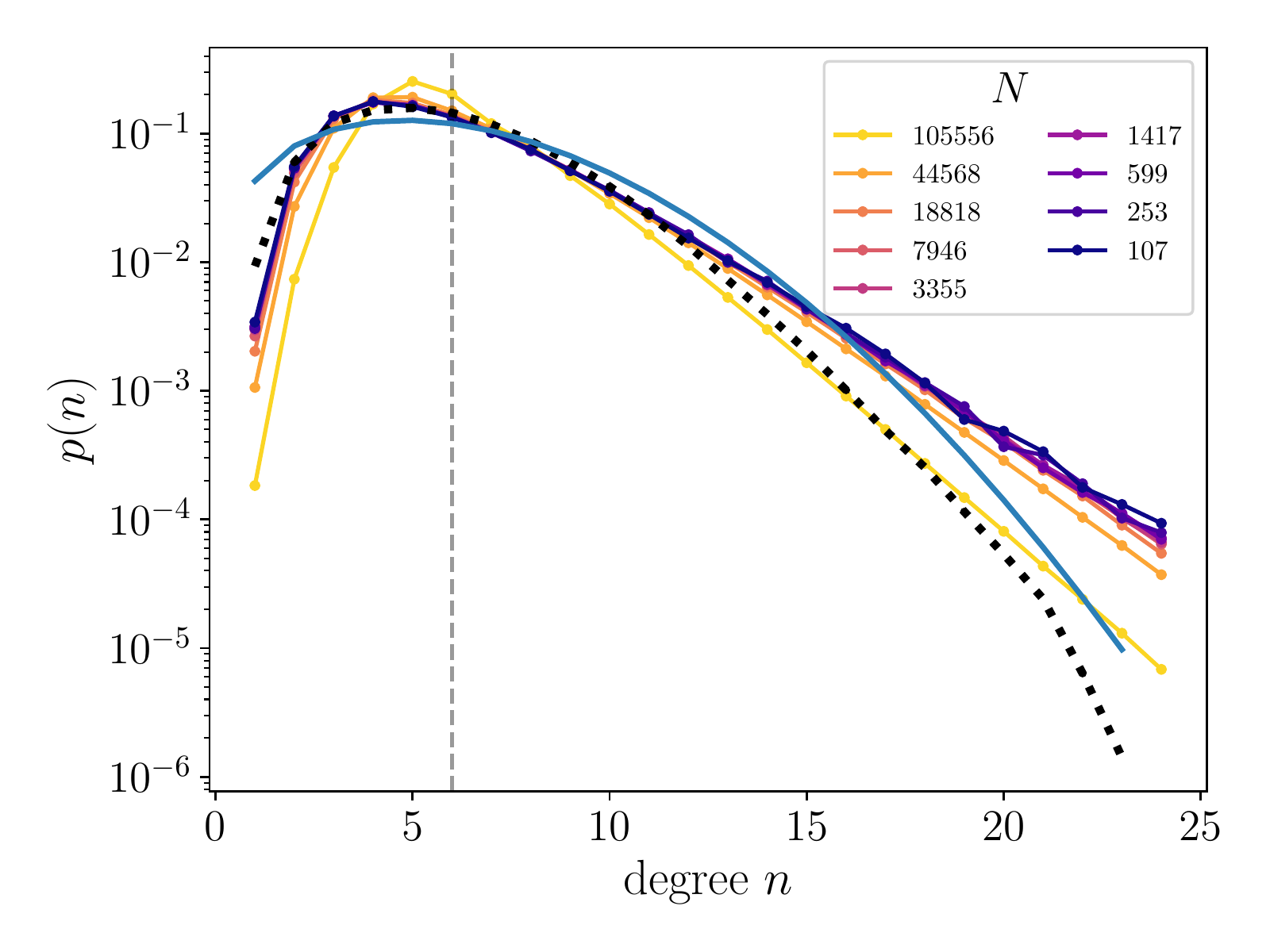}
    \includegraphics[width=.49\textwidth]{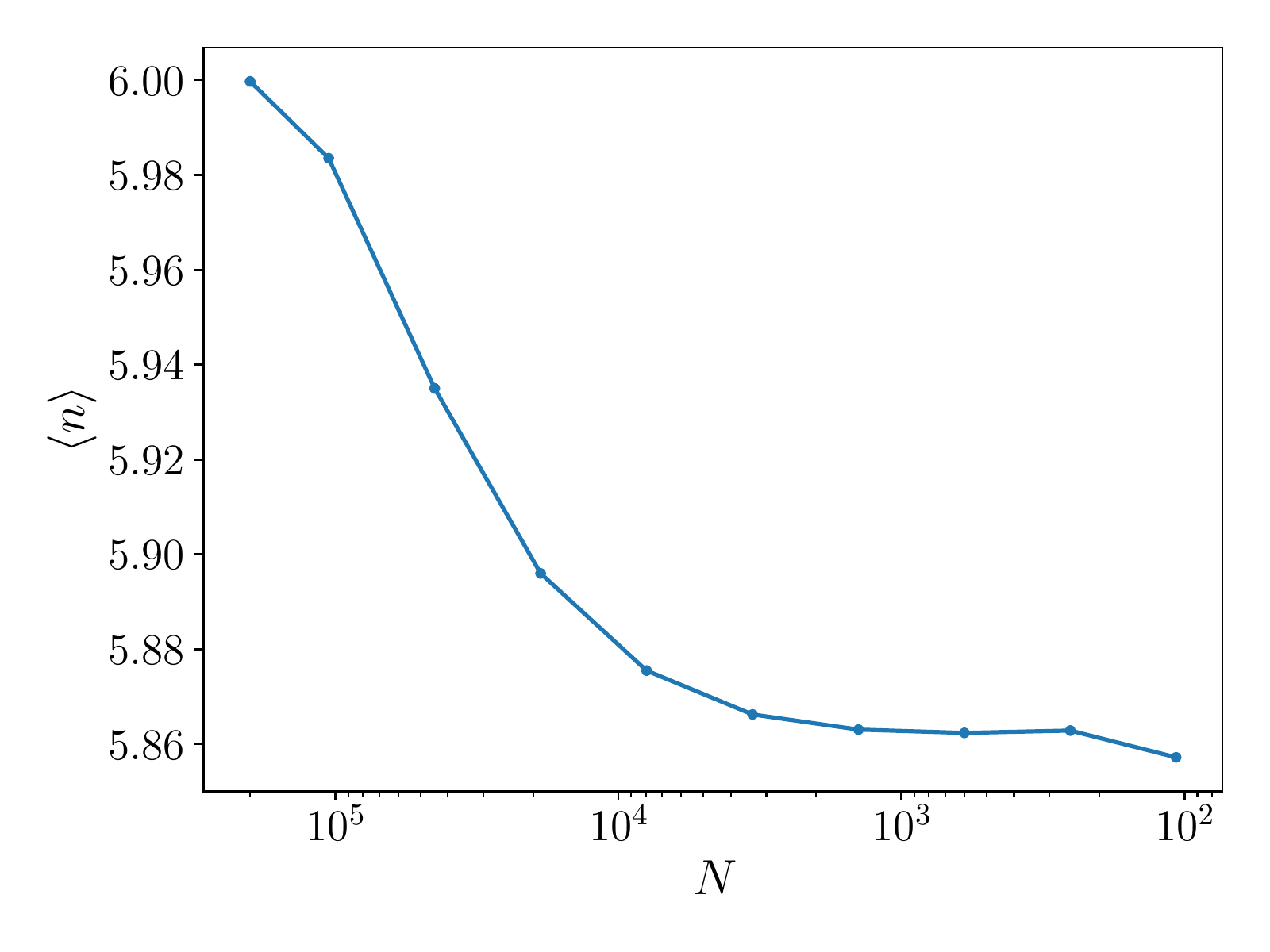}
    \caption{Degree distribution (left) and mean degree (right)  for various RG times in a geometrical RG which mimics some geometry-coupling correlations. The decimation rule repeatedly picks a spin with probability proportional to the inverse degree, and decimates a random edge adjacent to it. The blue line is the result of the analytical model with $w_n = 1/n$ from Eq.~\eqref{eq:gaussian_pn}, and the black dotted line is $p(n)$ from the fully correlated triangular RG at $N = 1417$  (as in Fig.~\ref{fig:degree_dist}). The mean degree drops slightly below $6$ and saturates.
    }
    \label{fig:degree_distributions_cull}
\end{figure}

We will take a simple physically motivated way of mimicking these observed correlations in the geometrical RG, in the hope that this brings us closer to the real two-dimensional fixed point. In order to select an edge to be contracted, we pick a spin $i$ with probability proportional to $1/n_i$ (where $n_i$ is the degree of spin $i$), and contract a random edge connected to it. This captures the phenomenon that higher-degree spins have weaker transverse fields and are adjacent to weaker bonds, and are therefore less likely to be involved in a decimation.

After running this geometrical RG we obtain a steady state $p(n)$ that is similar to what is seen in the correlated RG, as demonstrated in Fig.~\ref{fig:degree_distributions_cull}. Importantly the mean degree does not drop much below 6 which shows that the number of danglers being produced is very small, in line with the triangular RG. This qualitative agreement is evidence that the mechanism explained above is the one which produces the degree distribution in the scaling regime of the triangular RG.

\section{Correlationless RG equations for couplings}\label{app:couplingRG}

In this Appendix we provide details concerning the solution of Eqs.~\eqref{eq:secPequation} and \eqref{eq:secRequation}, reproduced here:
\begin{equation}\label{eq:appPequation}
    \frac{\partial P }{\partial \Gamma} = \frac{\partial P}{\partial \zeta} + \frac{R_0}{3} \sum_n p(n) \left[  -(n+2) P  + \sum_{i = 2}^n \int_0^\zeta d\zeta_1 P^{(n)}_{1i}(\zeta_1, \zeta - \zeta_1) \right]  + (P_0 +  R_0 ) P,
\end{equation}
\begin{equation}\label{eq:appRequation}
    \frac{\partial R}{\partial \Gamma} = \frac{\partial R}{\partial \beta}  + 3P_0 \int_0^\beta d\beta_1 \, R(\beta_1) R(\beta - \beta_1) + (-3P_0 + R_0)R.
\end{equation}

Here $P^{(n)}_{1i}$ is the joint distribution of the smallest and $i$th smallest variables when $n$ variables are independently drawn from the distribution $P$. Explicitly it is given by
\begin{equation}\label{eq:jointP}
    P^{(n)}_{1i}(\zeta_1 ,\zeta_i) = \Theta(\zeta_i - \zeta_1) \frac{n!}{(i-2)! (n-i)!} P(\zeta_1)P(\zeta_i) (C(\zeta_i)-C(\zeta_1))^{i-2} (1-C(\zeta_i))^{n-i}
\end{equation}
where $C$ is the cumulative distribution of $P$ given by $C(\zeta) = \int_0^{\zeta} d \zeta' \, P(\zeta').$ We shall normalize $\int_0^\infty P(\zeta) d\zeta=\int_0^\infty R(\beta) d\beta = 1$ at all RG times.

\subsection{Derivation}

Let us say the number of spins with log strength between $\beta$ and $\beta + d\beta$ is $S(\beta) d\beta$, so that $S(\beta) = N R(\beta)$. The analogous quantity for bonds is $B(\zeta) = 3N P(\zeta)$, since the number of bonds is $3N$. We will find the evolution under RG of the functions $B$ and $S$, and translate these to equations for $R$ and $P$ respectively. When $b$ bond decimations take place, the bond distribution evolves as 
\begin{equation}
    \delta B_b = \frac{b}{B_0} \frac{\partial B}{\partial \zeta} - 2 b P.
\end{equation}
The first term comes from the redefinition of $\zeta$ because the cutoff of the distribution is at a lower bare energy. The second term comes from the two additional bonds that are destroyed when a bond decimation occurs (within triangulations). The site distribution evolves as
\begin{equation}
    \delta S_b = -2b R + b \int d\beta_1 R(\beta_1) R(\beta-\beta_1),
\end{equation}
since every bond decimation destroys two log transverse fields and replaces them with their sum. 

If $s$ site decimations are performed, we similarly have
\begin{equation}
    \delta S_s = \frac{s}{S_0} \frac{\partial S }{\partial \beta }
\end{equation}
and
\begin{equation}
    \delta B_s = s\sum_n p(n) \left[  -(n+2) P  + \sum_{i = 2}^n \int_0^\zeta d\zeta_1 P^{(n)}_{1i}(\zeta_1, \zeta - \zeta_1) \right].
\end{equation}
Note that $\delta S = \delta(N R) \approx N \delta R + R\delta N  =  N \delta R - (s+b)R$. Similarly, $\delta B = 3N \delta P - 3(s+b) P$, so that 
\begin{equation}\label{eq:PRBSequations}
    \delta P =  \frac{1}{3N} \left(\delta B_s + \delta B_b + 3(s+b)P \right), \quad \delta R = \frac{1}{N} \left( \delta S_s + \delta S_b + (s+b)R \right).
\end{equation}
When the RG integrates out the log energy interval $\delta \Gamma$, the number of spin decimations is $s=S_0 \delta \Gamma= NR_0 \delta \Gamma$ and the number of bond decimations is $b =B_0 \delta \Gamma=  3N P_0 \delta \Gamma $. Plugging these into Eq.~\eqref{eq:PRBSequations} we obtain Eqs.~\eqref{eq:appPequation} and \eqref{eq:appRequation}.

\subsection{Scaling solution}

Given that we are looking for an IRFP, and we expect log energies to broaden in proportion to RG time $\Gamma$, we change variables to $z=\zeta/\Gamma , y=\beta/\Gamma $ and attempt scaling solutions 
\begin{equation}
    P(\zeta) = \frac{1}{\Gamma} \Pi\left(\frac{\zeta}{\Gamma}\right), \quad  
    R(\beta) = \frac{1}{\Gamma} \rho\left(\frac\beta\Gamma\right)\,
\end{equation}
to get the equivalent equations
\begin{equation}\label{eq:piequation}
    -\Pi -z\Pi' = \Pi' - \frac{\overline{n} \rho_0}{3}\Pi + \frac{\rho_0}{3} \sum_n p(n) \sum_{i = 2}^n \int_0^z dz_1 \, \Pi^{(n)}_{1i} (z_1, z-z_1) + (\Pi_0 + \frac{\rho_0}{3})\Pi,
\end{equation}
and
\begin{equation}\label{eq:rhoequation}
    -\rho -y\rho' = \rho'+ 3\Pi_0 \int_0^y dy_1 \, \rho(y_1) \rho(y-y_1) + (-3\Pi_0 + \rho_0) \rho.
\end{equation}
where $\Pi^{(n)}_{1i}$ is defined in analogy to $P^{(n)}_{1i}$ using $\Pi(z)$ instead of $P(\zeta)$, and $\overline{n} \equiv \langle n \rangle= \sum_n (np(n))$.

The RG numerics prompts us to try as a solution to Eq.~\eqref{eq:rhoequation} an exponential distribution $\rho = \rho_0 e^{-y\rho_0}$. Doing this yields   $\Pi_0 = 1/3$, regardless of $\rho_0$. We can turn our attention to Eq.~\eqref{eq:piequation} for $\Pi$, with the constraint that $\Pi(0) = 1/3.$

First note that the integral is from $0$ to $z$ and therefore must vanish at $z = 0$. This fixes
\begin{equation}\label{eq:fixrho0}
    \rho_0  = \frac{9\Pi'(0) + 4}{\overline{n} - 1}.
\end{equation}

\subsection{Poisson degree distribution}

With a suitable $p(n)$ it is possible to use the form of the joint distribution of order statistics in Eq.~\eqref{eq:jointP} to do the sum over $n$ in Eq.~\eqref{eq:piequation} explicitly.\footnote{We thank Nicholas O'Dea for pointing this trick out.} Consider the distribution $p(n) = e^{-\lambda} \lambda^n / n!$. This is in fact a remarkably good approximation to the numerically observed $p(n)$: see the green dotted curve in Fig.~\ref{fig:degree_dist}.

First write
\begin{equation}
     \Pi^{(n)}_{1i}(z_1 ,z-z_1) = \Theta(z-2z_1) \frac{n!}{(i-2)! (n-i)!} \Pi(z_1)\Pi(z-z_1) (\mathcal C(z-z_1)- \mathcal C(z_1))^{i-2} (1-\mathcal C(z-z_1))^{n-i}
\end{equation}
(where $\mathcal{C}(z) = \int_0^z dz_1 \Pi(z_1)$) as 
\begin{equation}
    \Pi^{(n)}_{1i}(z_1 ,z-z_1) = \Theta \Pi_1 \Pi_2 \frac{n!}{(i-2)! (n-i)!}  A^{i-2} B^{n-i},
\end{equation}
noting in particular that none of the bundled-up objects here have $n$ or $i$ dependence. Now take the quantity
\begin{equation}
    \sum_{n=2}^\infty e^{-\lambda} \frac{\lambda^n}{n!}  \sum_{i = 2}^n \int_0^z dz_1 \,  \Theta \Pi_1 \Pi_2 \frac{n!}{(i-2)! (n-i)!}  A^{i-2} B^{n-i}
\end{equation}
and reverse the order of sums, taking the integral all the way out, to get
\begin{equation}
    e^{-\lambda} \int_0^z dz_1 \,  \Theta \Pi_1 \Pi_2 \sum_{i = 2}^\infty \sum_{n=i}^\infty  \frac{\lambda^i}{(i-2)! }  A^{i-2} 
    \frac{\lambda^{n-i}}{(n-i)!} B^{n-i}.
\end{equation}
The $n$ sum is trivially $e^{\lambda B}$ and we are left with
\begin{equation}
e^{-\lambda} \int_0^z dz_1 \,  \Theta \Pi_1 \Pi_2 e^{\lambda B} \sum_{i = 2}^\infty \lambda^2\frac{\lambda^{i-2}}{(i-2)! }  A^{i-2} .
\end{equation}
After doing another exponential sum, we finally have
\begin{equation}
    e^{-\lambda}\lambda^2 \int_0^z dz_1 \, \Theta \Pi_1 \Pi_2 e^{\lambda (B + A)}  = 
   \lambda^2 \int_0^{z/2} dz_1  \,  \Pi(z_1)\Pi(z-z_1)  e^{ -\lambda \mathcal C(z_1)}.
\end{equation}

Let us make a couple of simple changes. First, since the Poisson distribution is normalized so as to have support on $0,1,2,\dots$, but the $n$ sum starts at $2$, we should change the normalization by a factor of $1/(1-e^{-\lambda}(1+\lambda))$. Also, since $\overline{n} = \lambda$ holds for the distribution that has support at $n = 0$ and $1$, this should also be modified to
\begin{equation}\label{eq:nbar}
    \overline{n} = \lambda \frac{1-e^{-\lambda}}{1-e^{-\lambda}(1+\lambda)}.
\end{equation}
(For the physically relevant case of $\overline{n} = 6$ these modifications are minuscule.)
Thus Eq.~\eqref{eq:piequation} becomes
\begin{equation}\label{eq:poissonpieq}
 \frac{(-4 + \rho_0 (\overline{n}-1))\Pi -3(z+1)\Pi'}{\rho_0}
   = \frac{ \lambda^2}{1-e^{-\lambda} (1+\lambda)} \int_0^{z/2} dz_1  \,  \Pi(z_1)\Pi(z-z_1)  e^{ -\lambda C(z_1)}.
\end{equation}
We will take $\overline{n} = 6$ henceforth.

It is worth stressing that this degree distribution does not represent the steady state of any decimation process on graphs that we know of, and has been chosen simply to make analytical progress. In fact, it cannot even quite be a degree distribution for a triangulation since it has support on $n = 2$. Nevertheless, the fact that it agrees with the degree distribution in the numerical RG means that using it was sensible at the level of our approximation, which cannot make use of information about the graph ensemble beyond $p(n)$ anyway. 

\subsection{Ansatz}

\begin{figure}
    \centering
    \includegraphics[width = 0.6\textwidth]{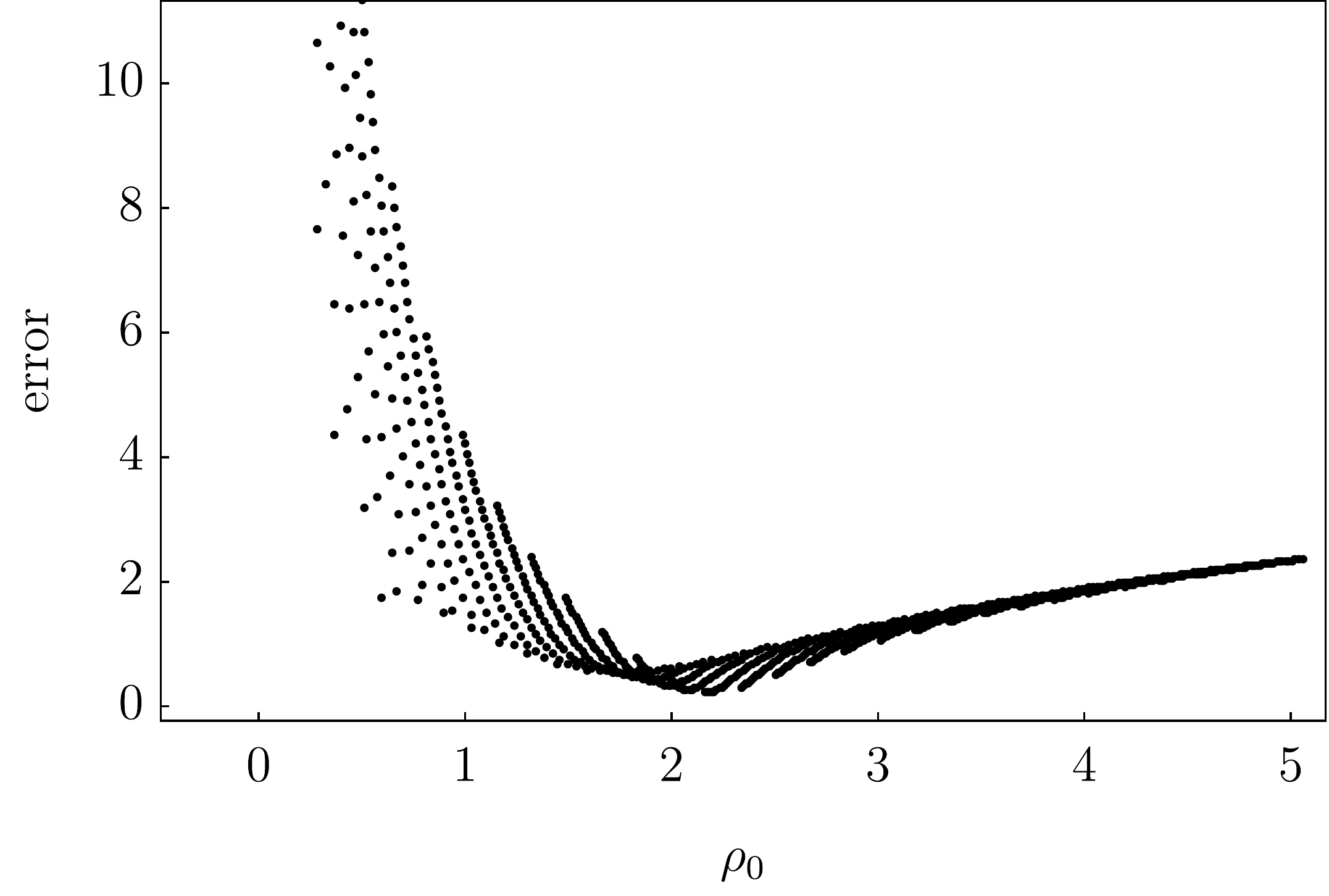}
    \caption{Error in the solution of Eq.~\eqref{eq:poissonpieq} versus the value of $\rho_0$, both evaluated by varying $c_1$ and $c_2$ through [0.0,3.0].
    The error is evaluated as $\sqrt{ \sum_{z \in \{0, \Delta z, 2\Delta z \dots, z_{max}\}} [ \mathcal L(z) -\mathcal  R(z)]^2  } $, where $\mathcal L(z)$ and $ \mathcal  R(z)$ are respectively the left- and right-hand sides of Eq.~\eqref{eq:poissonpieq}, and $z_{max} = 3 = 20 \times \Delta z$.
    }
    \label{fig:errrho0}
\end{figure}

Inspired by the shape of the fixed point $P(\zeta)$ we see in the numerics, we attempt the ansatz
\begin{equation}
    \Pi(z) = \left(\frac{1}{3}+c_1 z + c_2 z^2 \right)e^{-\gamma z}
\end{equation}
where $\gamma$ is fixed in terms of $c_1$ and $c_2$ by normalization. 

Also recall that demanding that the equation be satisfied at $z = 0$ fixes  $\rho_0$ 
 in terms of the parameters $c_1$ and $c_2$ which appear in $\Pi(z)$ (Eq.~\eqref{eq:fixrho0}). Thus we have two parameters which can be varied to look for the best approximation for a solution to Eq.~\eqref{eq:poissonpieq}. We did this and discovered that the error --- defined by the norm of a vector of differences between the two sides of Eq.~\eqref{eq:poissonpieq} sampled along a list of $z$'s  ---  is more or less a function of $\rho_0$ alone, as can be seen in Fig.~\ref{fig:errrho0}. The error is minimized by $\rho_0 \approx 2.2$.

\subsection{Critical exponents}

We can now find the critical exponent $\psi$. As discussed above, the number of spins reduced by going forward in the RG by log energy $d\Gamma$ is given by $(3P_0 + R_0)N d\Gamma$:
\begin{equation}
    \frac{dN}{d\Gamma} = -(3P_0 +   R_0) N = -(3\Pi_0 +  \rho_0) \frac{N}{\Gamma} \implies  N \propto \Gamma^{-(3\Pi_0 + \rho_0)},
\end{equation}
which gives $\psi = 2/(3\Pi_0 + \rho_0)$. Plugging in $\Pi_0 = 1/3$ and $\rho_0 \approx 2.2$ from above, we get $\psi \approx 0.6$. 

 The fractal dimension exponent $d_f$ can be obtained by including the auxiliary variable $\mu$ in Eq.~\eqref{eq:appRequation} and imposing that each bond decimation destroys two moments $\mu_1$ and $\mu_2$ of the spins being fused and creates the moment $\mu_1 + \mu_2$ (again, we assume no correlations). The resultant equation is identical to the 1d one~\cite{fisher1995critical}, and gives $\langle\mu\rangle \sim N^{-d_f/2} $, with $d_f = \psi (1 + \sqrt{1 + 4 \rho_0})/2 \approx 1.3.$

\bibliography{references}
\end{document}